
\font\ninerm=cmr9           \font\eightrm=cmr8 \font\sixrm=cmr6
\font\ninei=cmmi9           \font\eighti=cmmi8 \font\sixi=cmmi6
\font\ninesy=cmsy9          \font\eightsy=cmsy8 \font\sixsy=cmsy6
\font\ninebf=cmbx9          \font\eightbf=cmbx8 
\font\nineit=cmti9          \font\eightit=cmti8 \font\ninesl=cmsl9
\font\eightsl=cmsl8

 \font\twelvebf=cmbx12

\def\tenpoint{\def\rm{\fam0\tenrm}%

          \textfont0=\tenrm     \scriptfont0=\sevenrm
\scriptscriptfont0=\fiverm%
          \textfont1=\teni      \scriptfont1=\seveni
\scriptscriptfont1=\fivei%
          \textfont2=\tensy     \scriptfont2=\sevensy
\scriptscriptfont2=\fivesy%
          \textfont3=\tenex     \scriptfont3=\tenex
\scriptscriptfont3=\tenex%
          \textfont\itfam\tenit   \def\it{\fam\itfam\tenit}%
          \textfont\slfam\tensl   \def\sl{\fam\slfam\tensl}%
          \textfont\bffam\tenbf   \def\bf{\fam\bffam\tenbf}%
          \singleskip=12pt%
          \setbox\strutbox=\hbox{\vrule height8.5pt depth3.5pt width0pt}\rm}

\def\ninepoint{\def\rm{\fam0\ninerm}%
           \textfont0=\ninerm     \scriptfont0=\sixrm
\scriptscriptfont0=\fiverm%
           \textfont1=\ninei      \scriptfont1=\sixi
\scriptscriptfont1=\fivei%
           \textfont2=\ninesy     \scriptfont2=\sixsy
\scriptscriptfont2=\fivesy%
           \textfont3=\tenex      \scriptfont3=\tenex
\scriptscriptfont3=\tenex%
           \textfont\itfam\nineit   \def\it{\fam\itfam\nineit}%
           \textfont\slfam\ninesl   \def\sl{\fam\slfam\ninesl}%
           \textfont\bffam\ninebf   \def\bf{\fam\bffam\ninebf}%
           \singleskip=11pt%
           \setbox\strutbox=\hbox{\vrule height8pt depth3pt width0pt}%
           \rm}

\def\eightpoint{\def\rm{\fam0\eightrm}%
           \textfont0=\eightrm     \scriptfont0=\sixrm
\scriptscriptfont0=\fiverm%
           \textfont1=\eighti      \scriptfont1=\sixi
\scriptscriptfont1=\fivei%
           \textfont2=\eightsy     \scriptfont2=\sixsy
\scriptscriptfont2=\fivesy%
           \textfont3=\tenex       \scriptfont3=\tenex
\scriptscriptfont3=\tenex%
           \textfont\itfam\eightit   \def\it{\fam\itfam\eightit}%
           \textfont\slfam\eightsl   \def\sl{\fam\slfam\eightsl}%
           \textfont\bffam\eightbf   \def\bf{\fam\bffam\eightbf}%
           \singleskip=9pt%
           \setbox\strutbox=\hbox{\vrule height7pt depth2pt width0pt}%
           \rm}

\def\pagebreak{\par\vfill\supereject}

             \newif\ifnonum
\def\omitpagenumber{\global\nonumtrue}
\def\includepagenumber{\global\nonumfalse}
\def\nextpage{\pagebreak\includepagenumber}

             \footline={\ifnonum\nopagenumbers\else\hss\tenrm\folio\hss\fi}

             \newskip\singleskip            \singleskip=12pt plus 2pt
             \newskip\normalskip            \normalskip=18pt plus 2pt minus 2pt
             \newskip\doubleskip            \doubleskip=26pt plus 2pt minus 2pt

\def\singlespace{\baselineskip\singleskip}
\def\normalspace{\baselineskip\normalskip}

             \abovedisplayskip=12pt plus 3pt minus 1pt
             \belowdisplayskip=12pt plus 3pt minus 1pt
             \abovedisplayshortskip=5pt plus 3pt
             \belowdisplayshortskip=5pt plus 3pt

             \newcount\notenumber

\def\clearnotenumber{\global\notenumber=0}
           \clearnotenumber
\def\advancenote{\global\advance\notenumber by 1}
\def\note#1{\advancenote{\eightpoint\singlespace
                   {\footnote{$^{\the\notenumber}$}{#1}}}}

             \newbox\notebox

\def\endnotes{\setbox\notebox=\vbox{\centerline{\twelvebf Notes}\bigskip}

\global\def\note##1{\advancenote\setbox\notebox=\vbox{\unvbox\notebox

\bigskip\normalspace\item{\the\notenumber.}{##1\par}}$^{\the\notenumber}$}}

\def\symbolnote#1#2{{\eightpoint\singlespace{\footnote{$^{#1}$}{#2\smallskip}}}}

             \newcount\sectionnumber
             \newcount\subsectionnumber
             \newcount\subsubsectionnumber

\def\clearsectionnumber {\sectionnumber=0}
\def\clearsubsectionnumber {\subsectionnumber=0}
\def\clearsubsubsectionnumber {\subsubsectionnumber=0}

\clearsectionnumber \clearsubsectionnumber
\clearsubsubsectionnumber

\def\advancesection{\global\advance\sectionnumber by 1}
\def\advancesubsection{\global\advance\subsectionnumber by 1}
\def\advancesubsubsection{\global\advance\subsubsectionnumber by 1}

\outer\def\section#1{\advancesection \global\clearsubsectionnumber
           \global\clearsubsubsectionnumber
              \vskip0pt plus0.3\vsize\penalty-25\vskip0pt plus-0.3\vsize
              \bigskip\leftline
           {\twelvebf \the\sectionnumber .\hskip 12pt{#1}}
              \nobreak\medskip\vskip-\parskip}
\outer\def\subsection#1{\advancesubsection
\global\clearsubsubsectionnumber
              \medskip\leftline
           {\bf \the\sectionnumber .\the\subsectionnumber \hskip 12pt #1}
              \nobreak\smallskip\vskip-\parskip}
\outer\def\subsubsection#1{\advancesubsubsection
              \medskip
           {\it \the\sectionnumber .\the\subsectionnumber
.\the\subsubsectionnumber
            \hskip 12pt #1}
              \nobreak\smallskip\vskip-\parskip}

             \newcount\appendixnumber
             \newcount\subappendixnumber
             \newcount\subsubappendixnumber

\def\clearappendixnumber {\appendixnumber=0}
\def\clearsubappendixnumber {\subappendixnumber=0}
\def\clearsubsubappendixnumber {\subsubappendixnumber=0}

\clearappendixnumber \clearsubappendixnumber
\clearsubsubappendixnumber

\def\advanceappendix{\global\advance\appendixnumber by 1}
\def\advancesubappendix{\global\advance\subappendixnumber by 1}
\def\advancesubsubappendix{\global\advance\subsubappendixnumber by 1}

\outer\def\appendix#1{\advanceappendix\global\clearsubappendixnumber
           \global\clearsubsubappendixnumber\global\clearequationnumber
           \def\equation{\advanceequation\eqno(\hbox{\uppercase

\expandafter{\romannumeral\the\appendixnumber}.}\the\equationnumber)}
           \def\equate{\advanceequation&(\hbox{\uppercase

\expandafter{\romannumeral\the\appendixnumber}.}\the\equationnumber)}
           \pagebreak\leftline
           {\twelvebf Appendix
\uppercase\expandafter{\romannumeral\the\appendixnumber}.
               \hskip 12pt #1}
              \nobreak\medskip\vskip-\parskip}
\outer\def\subappendix#1{\advancesubappendix\global\clearsubsubappendixnumber
              \medskip\leftline
           {\bf
A\uppercase\expandafter{\romannumeral\the\appendixnumber}.\the\subappendixnumber
                \hskip 12pt #1}
              \nobreak\smallskip\vskip-\parskip}
\outer\def\subsubappendix#1{\advancesubsubappendix
              \medskip
           {\it
A\uppercase\expandafter{\romannumeral\the\appendixnumber}.\the\subappendixnumber
               .\the\subsubappendixnumber \hskip 12pt #1}
              \nobreak\smallskip\vskip-\parskip}

             \newcount\equationnumber

\def\clearequationnumber{\equationnumber=0}
           \clearequationnumber
\def\advanceequation{\global\advance\equationnumber by 1}
\def\equation{\advanceequation\eqno(\the\equationnumber)}
\def\equate{\advanceequation&(\the\equationnumber)}

             \newcount\figurenumber
             \newcount\tablenumber

\def\clearfigurenumber{\figurenumber=0}
           \clearfigurenumber
\def\advancefigure{\global\advance\figurenumber by 1}
\def\figurelong#1{\advancefigure\smallskip\centerline{\vbox{\narrower\narrower\e
ightpoint\singlespace
                      \noindent {\sl Figure \the\figurenumber .\quad#1}}}}
\def\figureshort#1{\advancefigure\smallskip\centerline{\eightpoint
                                 \sl Figure \the\figurenumber .\quad#1}}

\def\cleartablenumber{\tablenumber=0}
           \cleartablenumber
\def\advancetable{\global\advance\tablenumber by 1}
\def\tablelong#1{\advancetable\bigskip\centerline{\vbox{\narrower\narrower\eight
point\singlespace
                     \noindent {\sl Table \the\tablenumber .\quad#1}}}}
\def\tableshort#1{\advancetable\smallskip\centerline{\eightpoint
                                \sl Table \the\tablenumber .\quad#1}}

             \newcount\propositionnumber

\def\clearpropositionnumber{\propositionnumber=0}
           \clearpropositionnumber
\def\advanceproposition{\global\advance\propositionnumber by 1}
\def\proposition{Proposition \the\propositionnumber\advanceproposition}

             \newcount\definitionnumber

\def\cleardefinitionnumber{\definitionnumber=0}
           \cleardefinitionnumber
\def\advancedefinition{\global\advance\definitionnumber by 1}
\def\definition{\advancedefinition Definition \the\definitionnumber}

             \newcount\lemmanumber

\def\clearlemmanumber{\lemmanumber=0}
           \clearlemmanumber
\def\advancelemma{\global\advance\lemmanumber by 1}
\def\lemma{\advancelemma Lemma \the\lemmanumber}

\def\proclaim#1#2{\medskip\noindent{\bf #1:\quad\sl #2}}

\newbox\endproofsymbol
\setbox\endproofsymbol=\hbox{\hskip3pt\vrule width4pt height8pt
depth1.5pt} \outer\def\proof{\medskip\noindent{\bf
Proof:\enspace}}

\def\coco{\hbox{\tenrm coco$\,$}}
\def\conv{\hbox{\tenrm conv$\,$}}
\def\comp{\hbox{\tenrm comp$\,$}}

\def\and{\quad\hbox{and}\quad}
\def\blackboardR{\tenrm I\!\!R}

\def\frac#1#2{\hbox{$^{#1}\mskip -4mu/\!_{#2}$}}

\def\greatless{\,{\lower 6pt\hbox{$>$}\atop\raise 6pt\hbox{$<$}}\,}
\def\lessgreat{\,{\lower 6pt\hbox{$<$}\atop\raise 6pt\hbox{$>$}}\,}

\def\ref{\the}

             \newdimen\picwidth
             \newdimen\picheight
             \newdimen\difference
             \newdimen\movement

\def\picture (#1 scaled #2){\special{picture #1 scaled #2}}

\def\centerpicture (#1 scaled #2) (#3 by #4)%
           {\picwidth=#3
            \picheight=#4
            \divide\picwidth by 1000   \divide\picheight by 1000
            \multiply\picwidth by #2   \multiply\picheight by #2
            \advance\picheight by \abovedisplayskip
            \multiply\picwidth by -1
            \difference=\hsize
            \advance\difference by \picwidth
            \divide\difference by 2
            \movement=\difference
            \vskip \picheight
            \moveright\movement\vbox{\picture (#1 scaled #2)}}

\def\boxline#1#2{\vbox{\hrule
                            \hbox{\vrule\kern #1pt
                                \vbox{\kern #1pt\vbox{#2}\kern #1pt}
                                  \kern #1pt\vrule}
                               \hrule}}

\def\boxit#1#2{\vbox{\hrule
                            \hbox{\vrule\kern #1pt
                                \vbox{\kern #1pt\hbox{#2}\kern #1pt}
                                  \kern #1pt\vrule}
                             \hrule}}

\def\today{\ifcase\month \or January\or February\or March\or April\or
May\or June\or July\or August\or September\or October\or
November\or December\fi\space \number\year}

\def\qed{\vbox{\hrule\hbox{\vrule\kern3pt\vbox{\kern6pt}\kern3pt\vrule}\hrule}}

\def\qed{\vbox{\hrule\hbox{\vrule\kern3pt\vbox{\kern6pt}\kern3pt
\vrule}\hrule}}

\def\ao{\hbox{\sevenbf a}}
\def\aa{\hbox{\ninebf a}}
\def\po{\hbox{\sevenbf p}}

\def\BO{\hbox{\sevenbf B}}

\def\BB{\hbox{\ninebf B}}

\def\AO{\hbox{\sevenbf A}}
\def\AAA{\hbox{\tenbf A}}
\def\PPP{\hbox{\tenbf P}}

\def\pp{\hbox{\ninebf p}}

\def\e{{\cal E}}

\def\d{{\cal D}}

\def\b{{\cal B}}
\def\qed{\vbox{\hrule\hbox{\vrule\kern3pt\vbox{\kern6pt}\kern3pt\vrule}
\hrule}}

\normalspace
\pageno=1
\magnification=1200

\hfuzz=1pt \topinsert \vskip 1truein
\endinsert
\omitpagenumber
\normalspace
\centerline{}
\medskip

{\baselineskip=14pt\centerline{\twelvebf
Lindahl Equilibrium as a Collective Choice Rule\symbolnote{\dagger}{This research
was supported
by grants
from
the National Science Foundation.}}

\vskip .75truein \centerline{Faruk Gul and Wolfgang Pesendorfer}

\bigskip
\centerline{Princeton University} \vskip .5truein
\centerline{\today}

\bigskip

\centerline{\bf Abstract}

A collective choice problem is a finite set of social alternatives and a finite set of economic agents with vNM utility functions.  We associate a public goods economy with each collective choice problem and establish the existence and efficiency of (equal income) Lindahl equilibrium allocations. We interpret collective choice problems as cooperative bargaining problems and define a set-valued solution concept, {\it the equitable solution} (ES). We provide axioms that characterize ES and show that ES contains the Nash bargaining solution.  Our main result shows that the set of ES payoffs is the same a the set of Lindahl equilibrium payoffs. We consider two applications: in the first, we show that in a large class of matching problems without transfers the set of Lindahl equilibrium payoffs is the same as the set of (equal income) Walrasian equilibrium payoffs.  In our second application, we show that in any discrete exchange economy without transfers every Walrasian equilibrium payoff is a Lindahl equilibrium payoff of the corresponding collective choice market. Moreover, for any cooperative bargaining problem, it is possible to define a set of commodities so that the resulting economy's utility possibility set is that bargaining problem {\it and} the resulting economy's set of Walrasian equilibrium payoffs is the same as the set of Lindahl equilibrium payoffs of the corresponding collective choice market.

\bigskip

\medskip

\vfill \omitpagenumber \nextpage \pageno=1 \normalspace

\section{Introduction}

Consider an organization that must decide among several alternatives that affect the welfare of its members.  The goal is to find an equitable and efficient solution to the problem when no transfers among members are possible. Examples include a community's decision on how to allocate infrastructure investments among neighborhoods, the allocation of office space among groups within an organization, or the assignment of college roommates.

In this paper, we analyze the following mechanism.  Each group member is given an equal budget of fiat money and confronts a price for each of the relevant alternatives under consideration.  As in standard consumer theory, members choose an alternative that maximizes their utility subject to the budget constraint.  The organization acts as an auctioneer and implements an alternative that maximizes revenue.  We allow choices to be stochastic; that is, agents choose lotteries over social outcomes.  We also allow the prices for the various social outcomes to be individual-specific.  Thus, our mechanism is analogous to a Lindahl mechanism (Lindahl (1919)) in a public goods setting.  It differs from the standard Lindahl equilibrium in that the objects of choice are social outcomes (allocations) rather than individual consumptions.  We refer to this mechanism as a {\it collective choice market} and to its equilibria as Lindahl equilibria.

As an illustration, suppose a town with $n$ inhabitants considers implementing project A or project B.  The status quo yields zero utility for all inhabitants who are divided equally between A and B-supporters.  Each inhabitant receives utility $1$ if her favored alternative is implemented and zero otherwise.  In a collective choice market, every agent is given one unit of fiat money and prices for the two alternatives;  she must choose an optimal lottery over the two alternatives subject to her budget constraint.  In the Lindahl equilibrium,  B-supporters must pay $2$ for alternative B and zero for A;  A-supporters must pay $2$ for A and zero for B.  At these prices, it is optimal for all agents to choose the lottery that yields $A$ and $B$ with equal probabilities.  This lottery also maximizes auctioneer revenue and is thus a Lindahl equilibrium.

The town's decision problem can also be described as an n-person bargaining problem in which the attainable utility profiles correspond to the inhabitants' utilities for lotteries over $A$ and $B$.  More generally, every collective choice market can be mapped to an $n$-person bargaining problem.  We introduce a new solution concept for $n$-person bargaining problems - the {\it equitable solution}.  Unlike standard solution concepts such as the Nash bargaining solution or the Kalai-Smorodinsky solution, it is set valued. That is, it associates with every bargaining problem a set of equitable solutions.  Our main result (Theorem 2) relates Lindahl equilibrium outcomes and equitable outcomes of the associated bargaining game.

As the following example illustrates, it is often the case that several solutions to a bargaining game may qualify as equitable. Suppose Ann and Bob must divide two cakes, a peanut butter cake and a chocolate cake.  Utilities are linear but Bob is allergic to peanuts while Ann likes the peanut butter cake just as much as the chocolate cake.  One equitable division might be to give Ann the peanut butter cake and divide the chocolate cake equally between Ann and Bob.  To justify this outcome,  Ann could argue as follows:  {\it  Since Bob has no use for the peanut butter cake, from his perspective the situation is as if we only had the chocolate cake and in that situation it's obviously equitable to divide the chocolate cake equally.} Perles and Maschler (1981) provide an axiomatic foundation for Ann's argument.  The Perles-Maschler solution to the bargaining game would give Ann the peanut butter cake and divide the chocolate cake equally between the two players.  On the other hand, Bob could make the following argument: {\it If I were not allergic to peanuts we would each get one cake.  Since Ann is indifferent between the chocolate cake and the peanut butter cake it makes sense that she gets the peanut butter cake and I get the chocolate cake.} Nash (1951) provides an axiomatic foundation for Bob's argument.  In the Nash bargaining solution of this game, Ann gets the peanut butter cake and Bob gets the chocolate cake.  Of course, in-between solutions can also be justified.  For example, the Kalai-Smorodinsky solution suggests giving 2/3 of the chocolate cake to Bob and 1/3 to Ann while Ann gets all of the peanut butter cake.

To define the equitable solution, we first identify a subclass of bargaining problems that have a self-evidently fair outcome.  In those bargaining problems agents have linear utility over a fixed surplus to be divided among them and all standard bargaining solutions agree that equal division is the only plausible equitable outcome.\note{See Thomson (1994) for a survey of bargaining solutions.  For our setting, the relevant bargaining solutions are ordinal, that is, solutions that are invariant to affine transformation of utilities.  The Nash bargaining solution (Nash (1950)), the Kalai-Smorodinsky solution (Kalai and Smorodinsky (1975)) and the Perles-Maschler solution (Perles and Maschler (1981)) are examples of ordinal solutions.}  For a general bargaining problem; that is, when a fair solution is not self-evident, the equitable solution picks all those outcomes which are the obvious fair outcome of some dominating bargaining problem.\note{Bargaining problem $A$ dominates problem $B$ if for every utility profile in $B$ there is a better utility profile in $A$ and for every utility profile in $A$ there is a worse utility profile in $B$.}  In Theorem 1 we provide an axiomatic foundation for the equitable solution.

Our main result, Theorem 2, shows that the set of Lindahl equilibrium payoffs coincides with the equitable outcomes of the corresponding bargaining game.  One direction, every Lindahl equilibrium is equitable, can be interpreted as an equity-analogue of the first theorem of welfare economics.  It shows that our axioms capture the sense in which equilibrium outcomes of collective choice markets are equitable.  The second direction,  every equitable outcome can be achieved as a Lindahl equilibrium, is analogous to the second theorem of welfare economics.  It shows that collective choice markets are flexible enough to implement any outcome that is equitable.

We first apply our results to matching.  A group of individuals must decide who matches with whom as, for example, in a roommate assignment problem or in the classic two-sided matching problem (Gale and Shapley (1962)).  We show that in matching problems without transfers the Lindahl equilibrium outcomes of a collective choice market coincide with the outcomes of Walrasian equilibria in a market for partners.  In a collective choice market, consumers express their demands for social outcomes (that is, allocations) while in a Walrasian equilibrium, consumers express their demand for potential partners.  In both cases, utility is non-transferable and consumers have the same budget of fiat money.   Our result implies that the Walrasian equilibria of matching markets with non-transferable utility and equal budgets are equitable.

In our second application, we consider the allocation of non-divisible goods when consumers have identical budgets of fiat money.  Every Walrasian equilibrium of this exchange economy is a Lindahl equilibrium of the corresponding collective choice market. However, the set of Walrasian equilibria is sensitive to the exact specification of the traded goods (that is, how property rights are defined) while Lindahl equilibria are not. More precisely, the set of Lindahl equilibrium payoffs for two collective choice markets is the same whenever the two choice problems lead to the same bargaining set (i.e., the same set of feasible utility vectors). In contrast, the set of Walrasian equilibrium payoffs may be different for two different exchange economies that yield the same bargaining set. We show that every possible bargaining set can be ``commodified" as an exchange economy; that is, it is possible to define an exchange economy that yields the particular bargaining set as its set of feasible utility vectors. Moreover, this commodification can be done in a manner that renders the set of Walrasian equilibrium payoffs equal to the set of Lindahl equilibrium payoffs.  Finally, we show that for 2-person economies, a commodification that renders the set of Walrasian equilibrium payoff identical to the Lindahl equilibrium payoffs can be done with gross substitutes preferences.

\subsection{Related Literature}

Our paper is related to the extensive literature on axiomatic bargaining theory (see Thomson 1994) for a survey).  Our axiomatic treatment is related to Nash (1950) and we discuss this relationship in detail after the statement of Theorem 1 below.  For 2-person bargaining, our solution concept includes the Kalai-Smorodinsky solution (Kalai and Smorodinsky (1975)) and the Perles-Maschler solution (Perles and Maschler (1981)) in the set of equitable outcomes.

Hylland and Zeckhauser (1979) were the first to propose Walrasian equilibria as solutions to stochastic allocation problems.  Gul, Pesendorfer and Zhang (2020) extend Hylland and Zeckhauser from unit demand preferences to general gross-substitutes preferences.  Collective choice markets allow for arbitrary preferences, public goods and externalities and hence provide a further generalization of the environment considered in these papers.

Foley (1967), Schmeidler and Vind (1972) and Varian (1974) associate equity with envy-freeness.  Walrasian equilibria with equal budgets are envy free and, thus, these authors establish a connection between competitive outcomes and equity.  In a public goods setting, two agents may contribute different amounts to the same public good and hence it is not straightforward to adapt the notion of envy freeness to Lindahl equilibria.  Moreover, as we show in section 5.2,  there are multiple ways to commodify each collective choice market.  The same utility profile may be envy free for one specification but fail envy-freeness for another.  Thus, collective choice markets do not lend themselves to a coherent definition of envy-freeness. Instead, we provide a definition of equitable outcomes based on the associated cooperative bargaining problem and show its equivalence to Lindahl equilibria.\note{Sato (1987) adapts the notion of envy-freeness to Lindahl equilibria by assuming that agent $i$ converts $j$'s actual consumption of the public good into a {\it virtual quantity} based on $j$'s utility of that good. In effect, Sato identifies a commodity space and associated utility functions for which envy freeness of Lindahl equilibria with equal budgets is satisfied. 
}

\section{Collective Choice Markets and Lindahl Equilibrium}

Consider the following collective choice problem: $n$ agents, $i\in \{1,\dots, n\}$, must decide on one of $k$ social outcomes, $j\in K=\{1,\dots, k\}$. A random outcome, $q\in Q:=\{\hat q\in \blackboardR_+^k\,|\,\sum_{j\in K}\hat q^j=1\}$, is a probability distribution over the social outcomes. Agents are expected utility maximizers;  $i$'s utility if outcome $j\in K$ occurs is $u_i^j\geq 0$ and $u_i=(u_i^1,\ldots, u_i^k)$ denotes $i$'s utility index. In addition to the $k$ outcomes described above,  there is a disagreement outcome that yields zero utility to every agent. We dismiss all agents who have no stake in the collective decision; that is, we assume that for every utility $u_i$ there is some $j$ such that $u^j_i>0$. The vector $u=(u_1,\dots, u_n)$ denotes a profile of utilities.

We will define a social choice rule by identifying the Lindahl equilibria of a corresponding market economy, the {\it collective choice market}. This market has $n+1$ agents, the $n$ described above, now called consumers, and one firm. Consumer $i$ has one unit of fiat money and can purchase quantity $q^j\geq 0$ of each ``public good'' $j\in \{1,\ldots, k\}$ at price $p_i^j\ge 0$.  Outcome $0$ is identified with not purchasing any good (and therefore has price $0$). Let $e=(1,\ldots, 1)$ be the $k-$dimensional unit vector. Consumer $i$ faces prices $p_i=(p_i^1,\dots, p_i^k)$ and solves the following maximization problem:
$$U_i(p)=\max_{q} u_i\cdot q\hbox{ subject to }p_i \cdot q \leq 1, \,e\cdot q\leq 1\eqno(1)$$ We say that $q$ is an minimal-cost solution to the maximization problem above if $q$ solves that problem and $p\cdot \hat q\leq p\cdot q$ for every other solution $\hat q$.
The price paid for outcome $j$ depends on the identity of the consumer.  We let $p_i=(p_i^1,\ldots, p_i^k)\in \blackboardR_+^{kn}$ be consumer $i$'s prices and $p=(p_1,\ldots, p_n)$ be a price profile.  The firm chooses $q$ to maximize profit, that is, to solve
$$R(p)=\max_{q}\sum_{i=1}^np_i\cdot q \hbox{ subject to } e\cdot q =1\eqno(2)$$

\proclaim{Definition}{The pair $(p,q)$ is a {\it Lindahl equilibrium} (LE) for the collective choice market if $q$ is a minimal-cost solution  to every consumer's maximization problem at prices $p_i$ and solves the firm's maximization problem at prices $p$.}
\medskip

As is well known, a Lindahl equilibrium can be re-interpreted as a Walrasian equilibrium of a suitably modified alternative economy.  To do so, we identify $i$'s consumption of public good $j$ with a distinct private good $(i,j)$ and assume the firm can meet demand  $(q^{1j}.\ldots,q^{nj})$ with supply $q$ if $\max_{1\le i\le n} q^{ij}\leq q^j$ for all $j$.  This connection to Walrasian equilibrium means that existence of a Lindahl equilibrium can be shown by adapting standard results.  The proof of Lemma 1 and subsequent results are in the appendix.
\smallskip

\proclaim{Lemma 1}{Every collective choice market has a Lindahl equilibrium; all Lindahl equilibria are Pareto efficient.}

\medskip
A broad range of applications including all discrete allocation and matching problems can be modeled as collective choice markets. For example, if the aggregate endowment consists of a collection of indivisible goods, then the set of outcomes is simply the set of all allocations.

Similarly, to map a two-sided matching problem into a collective choice market, we partition ``consumers'' into two groups, $N_1, N_2$ and let the outcomes be the set of all matchings; that is, one-to-one functions $j:N_1\cup N_2\rightarrow N_1\cup N_2$ such that $j(j(i))=i$ and [$i\in N_l$ implies $j(i)=i$ or $j(i)\notin N_l$]. Hence,  a member of group $l$ is either unmatched ($j(i)=i$) or is matched with someone from the other group ($j(i)\notin N_l$).  In section 4, we analyze these applications and relate Lindahl equilibrium outcomes to standard Walrasian outcomes of these economies.

Let $L(u)\subset\blackboardR_+^n$ be the set of utility profiles that can be supported as a LE of the corresponding collective choice market $u$; that is, $L(u)=\{(u_1\cdot q,\dots,u_n\cdot q)\,|\,(p,q)\hbox{ is a LE of }u\}$.

\section{The Bargaining Problem}

In this section, we define and characterize the equitable solution, a cooperative solution concept. For any $x, y\in\blackboardR^n$, we write $x\leq y$ to mean $x_i\leq y_i$ all $i$. For any bounded set $X$, we let $d(X)$ (the {\it disagreement point} of $X$) denote the infimum of $X$ in $\blackboardR^n$; that is, (i) $d(X)\leq x$ for all $x\in X$ and (ii) $y\leq x$ for all $x\in B$ implies $y\leq d(X)$. Similarly, we let $b(X)$ (the {\it bliss point} $X$) denote the supremum of any bounded $X$; that is, $b(B)=-d(-B)$.

For any finite, bounded set $X\subset\blackboardR^n$, let $\conv X$ denote its convex hull and $\comp X$ denote its comprehensive hull of $X$; that is, $$\comp X:=\{x\in\blackboardR^n\,|\, d(X)\leq x\leq y\hbox{ for some }y\in X\}$$ Then, the set $\coco X:=\comp\conv X$ is it convex comprehensive hull; that is, the smallest (in terms of set inclusion) convex and comprehensive set that contains $X$.

For any $a\in\blackboardR^n$ and $x$, let $a\otimes x=(a_1\cdot z_1,\dots, a_n\cdot z_n)$, $a\otimes B=\{a\otimes x\,|\,x\in B\}$ and $B+z=\{x+z\,|\,x\in B\}$. Let $e^i$ denote unit vector with zeros in every coordinate except $i$,  $o:=(0,\dots, 0)\in\blackboardR^n$, $e:=(1,1,\dots ,1)\in\blackboardR^n$ and $\Delta=\conv\{o,e^1,\dots, e^n\}$.

We consider bargaining problems in which the set of attainable utility profiles is a full dimensional, comprehensive polytope. Let $\b$ denote the set of all such polytopes and $\b_o=\{B\in\b\,|\,o=d(B)\}$ denote the set of all polytopes with disagreement point $o$; that is, the set of normalized polytopes. A (set-valued) bargaining solution is a mapping $S:\b\rightarrow2^{\blackboardR^n}\backslash\emptyset$ such that $S(B)\subset B$.  Hence, a bargaining solution chooses a nonempty set of alternatives from every bargaining problem. Below, we define a set-valued bargaining solution, which we call the {\it equitable solution} (ES) and provide axioms that characterize it.

Equal division is the natural fair outcome if the bargaining set is the unit simplex.  Since von Neumann-Morgenstern utilities are unique only up to positive affine transformations, any affine transformation of the unit simplex should yield the corresponding affine transformation of the fair outcome.  Let $\d$ be the bargaining problems that are positive affine transformations of the unit simplex, that is,  $B\in\d\subset \b$ if  there exists $a,z\in \blackboardR^n$ such that $a_i>0$ for all $i$ and $B=a\otimes \Delta+z$.  For $B=a\otimes \Delta+z\in \d$ the {\it fair} outcome is $x={1\over n}a\otimes e+z$.  Hence, only bargaining problems that are equivalent, up to a positive affine transformation of utilities, to $\Delta$ have fair outcomes, and the fair outcome is the corresponding affine transformation of to the unique symmetric and efficient outcome of $\Delta$. Let $F(B)$ be the set of fair outcomes of $B\in \b$ with the convention that $F(B)=\emptyset$ if $B\not\in \d$.

\proclaim{Definition}{Let $A,B\in\b$. Then, $A\geq B$ if for every $x\in A, y\in B$, there exist $x'\in A$, $y'\in B$ such that $x'\geq y$ and $x\geq y'$.}

\medskip

Thus, $A\ge B$ if for every utility vector is in $B$ there is a corresponding utility vector in $A$ that dominates it and, conversely, for every utility vector in $A$ there is a corresponding utility vector in $B$ that is dominated by it.
The equitable solution consists of those outcomes of $B$ that coincide with the fair outcome of a simplex $A\ge B$.

\proclaim{Definition}{The equitable solution is $E(B):=\{x\in B\cap F(A)\,|\,A\geq B\}$.}

\medskip
With some abuse of terminology, we also refer to the mapping from bargaining problems to their equitable sets as ES. Below, we provide axioms on the bargaining solution $S$ that characterize the equitable solution. The first axiom,  scale-invariance, is  familiar from other bargaining solutions including the Nash bargaining solution and the Kalai-Smorodinsky solution. It asserts that positive affine transformations of utilities do not change the set of chosen (physical) outcomes.

\proclaim{Scale Invariance}{$S(a \otimes B+z)=a\otimes S(B)+z$ whenever $a_i>0$ for all $i$.}

Our symmetry axiom applies only to the bargaining problem $\Delta$. It ensures that the unique symmetric and efficient outcome of $\Delta$ is the only one chosen.

\proclaim{Symmetry}{$S(\Delta)=\{{1\over n}\cdot e\}$.}

The following axiom is similar to independence of irrelevant alternatives (IIA) of the Nash bargaining solution.

\proclaim{Consistency}{$B\leq A$ implies $S(A)\cap B\subset S(B)$.}

\smallskip
One important difference between consistency and IIA is that the former is applicable even if $d(A)\not=d(B)$. The latter replaces $\leq$ with $\subset$ and requires $d(A)=d(B)$. The next axiom, justifiability, is our main assumption.  We discuss its interpretation following Theorem 1 below.

\proclaim{Justifiability}{$x\in S(B)$ implies $B\leq A$ and $\{x\}=S(A)$ for some $A$.}

\medskip
Theorem 1 below reveals that the preceding four axioms characterize ES.

\proclaim{Theorem 1}{The only bargaining solution that satisfies the four axioms above is ES.}

\smallskip
It is easy to verify that $E$ satisfies the axioms. For the converse, let $S$ be a bargaining solution that satisfies the axioms and note that scale-invariance and symmetry imply that $S(a\otimes \Delta+z)={1\over n}\{a\otimes e+z\}=F(a\otimes \Delta+z)$ for all $a\in\blackboardR^n_{++}$ and $z$. Then, consistency, ensures $E(B)\subset S(B)$ while justifiability implies $x\notin S(B)$ whenever $x\notin S(B)$ for all $B$.

To see how our axioms relate to the axioms for the Nash bargaining solution (Nash 1950), consider a single-valued solution. In that case, adding $d(A)=d(B)$ to consistency and justifiability implies that the Nash bargaining solution is characterized by either of these two axioms together with scale invariance and symmetry. We can get this result even if we replace $\leq$ with $\subset$  while maintaining $d(A)=d(B)$ in both consistency and justifiability.

The equitable solution is a permissive solution concept that, in the case of two agents, includes all the standard scale invariant bargaining solutions.  In particular, the Nash bargaining solution (Nash (1950)), the Kalai-Smorodinsky solution (Kalai and Smorodinsky (1975)) and the Perles-Maschler solution (Perles and Maschler (1981)) are contained in the equitable solution when $n=2$.

To see the motivation for  justifiability consider again the example in the introduction:  Ann and Bob must divide two cakes, a peanut butter cake and a chocolate cake.  Utilities are linear but Bob is allergic to peanuts while Ann likes the peanut butter cake just as much as the chocolate cake.  The corresponding bargaining set $B$ is the convex hull of the points $\{(0,0), (1,0), (1/2,1), (0,1)\}$.  Ann could argue as follows:  {\it I should get the peanut butter cake since you don't have any use for it.  We both like the chocolate cake and so it makes sense to divide it equally.  So, we should choose $x=(3/4,1/2)$ as our solution.} Ann's argument leads to the Perles-Maschler solution of this bargaining game.  On the other hand, Bob could make the following argument: {\it If I were not allergic to peanuts we would each get one cake.  Since you are indifferent between the chocolate and the peanut butter cakes it makes sense that you get the peanut butter cake and I get the chocolate cake.  So we should choose $y=(1/2,1)$ as our solution.} Bob's argument leads to the Nash-bargaining solution of this bargaining game.

Justifiability says that both Ann and Bob have valid arguments and, thus, permits both solutions as possible outcomes.  More precisely, Ann can justify her favored outcome by noting that the bargaining game $A=\hbox{conv}\{(1/2,0),(1,0),(1/2,1)\}$ is a translation of a simplex and its fair outcome is $x$.  Since $A\ge B$, her proposed alternative $x$ is justified.  Bob can justify his favored outcome by pointing to the bargaining game $A'=\hbox{conv}\{(0,0), (1,0), (0,2)\}\ge B$ which has $y$ as its fair outcome.

The following lemma characterizes the equitable solution for the 2-player case and yields as a straightforward corollary the fact that the three solutions above are contained in the equitable solution.

\proclaim{Lemma 2}{If $n=2$, then $E(B)$ consists of all Pareto efficient utility profiles $x\in B$ such that $x\ge d(B)+(b(B)-d(B))/2$.}

\medskip
Lemma 2 shows that in the case of two agents the equitable solution consists of all Pareto efficient outcomes that satisfy {\it midpoint domination}.  The midpoint corresponds to the outcome that would be achieved if a coin is flipped to decide which agent is made a dictator.  In his survey of bargaining solutions, Thomson (1994, p. 1254) points out that most bargaining solutions satisfy midpoint domination.  In particular, the three solutions mentioned above satisfy it.

As Lemma 2 shows, the two agent case allows a simple characterization of the set of equitable outcomes.  However, when there are more than 2 agents this is no longer the case.  In particular, while the equitable solution continues to satisfy midpoint domination, the converse is no longer true.  Not every Pareto efficient outcome that satisfies midpoint domination can be equitable.   For example consider the 3-person bargaining problem $B=\hbox{coco}\{(0,0,0), (1,1,1/2), (1/3,1/3,1)\}$.  Then, $x=(1/3,1/3,1)$ is a Pareto efficient outcome that satisfies midpoint domination (since each agent receives at least $1/3$) but it is not an equitable solution since there is no $A\in \d, A\ge B$ that has $x$ as its fair outcome.  Our main result, Theorem 2 below, characterizes the set of equitable outcomes as the set of Lindahl equilibrium outcomes of the corresponding collective choice market.

Before doing so, we relate the equitable solution to the Nash bargaining solution.  For any $B\in \b$, let $$f_B(x):=\sum_i\log(x_i-d_i(B))$$
The Nash bargaining solution of $B$, $\eta(B)$,  is the unique outcome that maximizes $f_B(\cdot)$ in $B$.

\proclaim{Definition}{The outcome $x\in B$ is Nash-sustainable if there exists $A\in \b, A\ge B$ such that $x=\eta(A)$.}

\medskip
Note that the fair outcome coincides with the Nash bargaining solution if the bargaining set is a positive affine transformation of a simplex.  Therefore, any outcome in the equitable solution must be Nash sustainable.   Lemma 3, below, shows that the converse is true as well.  We let $N(B)$ denote the set of Nash-sustainable outcomes of the bargaining problem $B$.

\proclaim{Lemma 3}{For all $B\in\b$, $E(B)=N(B)$.}

\medskip
Lemma 3 shows that Nash sustainability is an alternative characterization of the outcome function that satisfies our axioms.

\section{The Equitable Solution and Lindahl Equilibria}

Any collective choice problem $u$, corresponds to a unique bargaining problem $$B_u:=\coco(\{(u_1^j,\dots, u_n^j)\,|\,j\in K\}\cup\{o\})$$  Hence, $B_u$ is the set of feasible utilities in the collective choice problem. Theorem 2 below is our main result. It shows that the equitable solution coincides with the Lindahl equilibrium payoffs.

One direction, every Lindahl equilibrium is equitable, shows that the equitable solution captures the notion of fairness embodied in Lindahl equilibria of collective choice markets.  Just as the first welfare theorem clarifies the notion of efficiency embodied in competitive models, Theorem 2 shows that Lindahl equilibria with equal budgets represent outcomes that are {\it justifiable} in the sense of our final axiom above.  The other direction, every equitable solution is a Lindahl equilibrium, shows that all equitable solutions can be implemented via a collective choice market.  Thus, this direction of Theorem 2 is analogous to the second welfare theorem albeit without the need to vary consumer's wealth.

\proclaim{Theorem 2}{The set of Lindahl equilibrium payoffs is the same as the equitable solution of the corresponding bargaining problem: $L(u)=E(B_u)$ for all $u$.}

\smallskip

To see how every equitable outcome can be made into a Lindahl equilibrium payoff, take any equitable $a\in B_u$. By definition, there exists some bargaining set $A$ that (i) is a positive affine transformation of a simplex; (ii) has $a$ as its fair outcome;  and (iii) satisfies $A\ge B_u$.  Clearly, $a$ must be on the Pareto-frontier of $B_{u}$.  In general, $d(A)$ need not be the origin but in the special case where it is we have $B_u\subset A$ and $A= na\otimes \Delta$.  Consider that special case. Since $a$ is Pareto efficient, there is a distribution $q$ over Pareto efficient outcomes that delivers the utility vector $a$.  The vector $q$ is the Lindahl equilibrium allocation. To find the Lindahl equilibrium prices, let $y^j=(u_1^j,\dots,u_n^j)$ be the utility vector corresponding to outcome $j$.  Since $B_u\subset A$ the vector $y$ must be a convex combination of the extreme points of $A=na\otimes \Delta$.  Hence, there are $z_i^j\ge 0$ such that $\sum_iz_i^j=1$ and $\sum_ina_iz_i^j\cdot e^i=y^j$ for all $j$. Then, set $p^j_i=nz_i^j$ to be the price agent $i$ pays for outcome $j$. It is then not difficult to verify that $(p,q)$ is a Lindahl equilibrium. Our proof that $E(B_u)\subset L(u)$ for all $u$ generalizes this argument to cover the case in which $d(A)\not=o$ when we have $B_{u}\leq A$ but not $B_{u}\subset A$.

For the converse; that is, to see that every Lindahl equilibrium payoff is equitable, let $(p,q)$ be a Lindahl equilibrium and recall that $q$ solves the linear program below for every consumer $i$:
$$\max_{q'} u_i\cdot q'\hbox{ subject to }p_i \cdot q' \leq 1, \,e\cdot q'\leq 1\eqno(1) $$
Let $\alpha_i$ be the shadow price of the constraint $p_i\cdot q'\leq 1$ and let $c_i$ be the shadow price of the constraint $e\cdot q'\leq 1$.  Then, the constraint of the dual of the above linear program requires that for all $i,j$,
$$\alpha_ip_i^j\geq u^j_i-c_i\eqno(3)$$
Moreover, optimality of $q$ implies that inequality (3) holds with equality if $q^j>0$.  Suppose $\alpha_i$, the shadow price of the budget constraint, is strictly positive for every consumer.  Then, $p_i\cdot q=1$ and, since $\alpha_ip_i^j=u^j_i-c_i$ if $q^j>0$, we have $\alpha_i+c_i=u_i\cdot q$ for all $i$; that is, $\alpha+c$, where $\alpha=(\alpha_1,\dots,\alpha_n)$ and $c=(c_1,\dots,c_n)$, is the payoff vector associated with the equilibrium $(p,q)$.   Note that $\alpha+c$ is the fair outcome of the bargaining game $n\alpha\otimes\Delta+c$. The key step of the proof establishes that $B_u\leq n\alpha\otimes\Delta+c$  which then implies that  $\alpha+c$ is an equitable outcome of $B_u$.

Theorem 2 shows that Lindahl equilibria are equitable, which, by Lemma 3, implies that every Lindahl equilibrium is Nash sustainable.  Thus, if $B_u$ is the bargaining game corresponding to the collective choice problem then each Lindahl equilibrium payoff is the payoff in the Nash bargaining solution of a bargaining game $A$ such that $A\ge B_u$.  In Corollary 1, below, we rely on this connection to characterize all Lindahl equilibrium allocations in terms of the Nash bargaining solution.

For any collective choice market $u$, the allocation $q$ that yields the Nash bargaining outcome is the unique solution to the maximization problem:
$$N^*=\max_{q\in Q}\,\sum_i\log (u_i\cdot q)\eqno(N)$$
We refer to the solution of (N) as the Nash allocation for $u$. Note that if $q$ is the Nash allocation of $u$, then  $N(B_u)=u\cdot q=(u_1\cdot q,\ldots, u_n\cdot q)$.

To relate Lindahl equilibria to Nash allocations, we define a family of utilities derived from the original utility $u$ of the collective choice problem.  Let $C_i(u_i)=[0,\max_j u_i^j)$ and $C(u)=\prod_{i=1}^n C_i(u_i)$. For $c_i\in C_i(u_i)$, define $$\bar u^j_i(c_i):=\max\{u^j_i-c_i,0\}$$
Since $c\in C(u)$, $\bar u^j_i(c_i)>0$ for some $j$ and hence $\bar u_i(c_i)=(u^1_i,\dots, u_i^k)$ is a utility. Let $\bar u(c)=(\bar u_1(c_1),\ldots, \bar u_n(c_n))$ be the corresponding profile of utilities.

\proclaim{Definition}{Let $q$ be the Nash allocation for $\bar u(c)$ for some $c\in C(u)$. Then, $c$ is admissible if $u_i^j \ge c_i$  for all $j$ such that $q^j>0$.}
\medskip

Thus, a vector $c$ is admissible if the corresponding Nash allocation only chooses alternatives that give each consumer a utility no less than $c_i$  (according to the original utility $u_i$).

\proclaim{Corollary 1}{If $q$ is the Nash allocation for $\bar u(c)$ and $c$ is admissible, then $(p,q)$ such that $p_i=\bar u_i(c_i)/(\bar u_i(c_i)\cdot q)$ is a Lindahl equilibrium for $u$. Conversely, if $q$ is a Lindahl allocation for $u$, then there is an admissible $c$ such that $q$ is a Nash allocation for $\bar u(c)$.}

\medskip
Corollary 1 shows that every Lindahl equilibrium allocation corresponds to the Nash allocation for utility $\bar u(c)$ for some admissible $c$.   To understand the need for admissibility, let $q$ be a solution to (N) for an arbitrary $c\in C(u)$.  For $q$ to be a Lindahl equilibrium, the payoff of agent $i$ must be $\bar u_i(c)\cdot q+c_i$;  note that $\bar u_i(c)\cdot q+c_i\ge u_i\cdot q$ but feasibility requires that this inequality holds with equality.   Equality holds if and only if $c$ is admissible.

Above we have interpreted $c_i$ as a parameter that modifies consumer $i$'s utility.  Alternatively, we can interpret the $c_i$'s as modifying the disagreement points in the bargaining game $B_u$.  The original disagreement point of $B_u$ is the origin since we assume all consumers get utility zero if no agreement is reached.  For $c\ge 0$, let $B_{u}(c):=\{u\in B_u\,| u_i\ge c_i\}$ be the bargaining game $B_u$ modified so that the disagreement point is $c$ instead.   We can re-state the definition of admissibility, above, to say that $c$ is admissible if the Nash bargaining solution of $B_u(c)$ can be implemented with an allocation that only puts weight on outcomes that give utility at least $c_i$ to consumer $i$.  Corollary 1 shows that every Lindahl equilibrium corresponds to a Nash bargaining solution with an admissible disagreement point $c$ and, conversely, every Nash bargaining solution with an admissible disagreement point $c$ can be implemented as a Lindahl equilibrium of the collective choice market.

\section{Applications}

In this section, we analyze two applications of the Lindahl collective choice rule.  In the first, we consider one-to-one matching problems without transfers;   in the second, we consider general allocation problems with indivisible goods and no transfers.

\subsection{Matching}

A group of agents must decide who matches with whom as, for example, in the problem of choosing roommates.  A matching, is a bijection $j$ from the set of all agents to itself such that $j(j(i))=i$ for all $i$. If $j(i)=i$, then $i$ is said to be unmatched. Some matchings may be infeasible. For example, agents could be workers and firms and firms matching with another firms may be disallowed.  The set $K=\{1,\dots, k\}$ represents the feasible matchings and let $N_i=\{m |\, j(i)=m, \hbox{ for some } j\in K\}$ be the feasible matches of agent $i$ and assume that $N_i\not=\emptyset$ for all $i$.

Let $w_i^m$ be the utility of agent $i$ when she matches with agent $m$. We normalize agents' utilities so that being unmatched yields 0 utility for every agent and assume that $w_i^m>0$ for some $m\in N_i$. We eliminate all matchings that are not individually rational and therefore, we assume $w_i^{j(i)}\geq 0$ for all $i$ and $j$. For notational convenience, we set $w_i^m=0$ if $m\not\in N_i$.

We identify the following competitive market with this matching problem: each agent has one unit of fiat money and agent $i$ must pay price $\pi_i^m$ for matching with agent $m$. The price of remaining unmatched $\pi_i^i$ is zero for all agents.  For notational convenience, we also set $\pi_i^m=0$ for $m\not\in N_i$.  Let  $\pi_i=(\pi_i^1,\dots,\pi_i^n)$ be the corresponding price vector and let $\pi=(\pi_1,\dots, \pi_n)$. A commodity, in this setting, is an ordered pair $(i,m)$ denoting $i$'s match with $m$. Since we allow randomization, $i$'s consumption of $(i,m)$ can vary between $0$ and $1$. Thus, $i$'s consumption set is: $$Z_i=\Big\{\xi_i\in\blackboardR_+^n\,\Big|\,\xi_i\cdot e\le 1\hbox{ and }\xi_i^m>0\hbox{ implies } m\in N_i\Big\}$$ Then, given prices $\pi$, each agents solves the following maximization problem:
$$W_i(p)=\max_{\xi_i\in Z_i}w_i\cdot \xi_i \hbox{ }\hbox{ subject to }\pi_i\cdot \xi_i \leq 1\eqno(\hbox{M1})$$ The consumption $\xi_i$ is a minimum cost solution to the consumer's problem if it solves the above maximization problem and no other solution costs less than $\xi_i$ at prices $\pi_i$.

From the firm's perspective, the revenue of matching $j$ is the sum of the prices of the individual matches; that is, $\sum_{i=1}^n\pi_i^{j(i)}$.  Therefore, the firm solves the maximization problem: $$R(\pi)=\max_{q\ge 0}\sum_{j\in K}\sum_{i=1}^n\pi_i^{j(i)}\cdot q^j\hbox{ subject to }q\cdot e \leq 1\eqno(\hbox{M2})$$
Let $\xi=(\xi_1,\ldots, \xi_n)$ be a vector of consumer choices. The triple $(\pi,\xi,q)$ is a (strong) Walrasian equilibrium of the matching market $v$ if $\xi_i$ is a minimum cost solution to consumer $i$'s problem (M1), $q$ solves the firm's maximization problem (M2) and for all $i,m$
$$\xi_i^m=\sum_{\{j|\,j(i)=m\}}q^j \eqno(\hbox{M3})$$
Equation (M3) is the market clearing condition; that is, the requirement that consumer $i$'s demand for matches with $m$ are met by the firm's supply of such matches.

The competitive economy differs from the collective choice market in the way consumers express their demands.  In the competitive economy, consumers specify demands for private goods (partners $m\in N_i$) while in a collective choice market,  consumers specify their desired collective goods (matches $j\in K$).   To map the matching market into a collective choice market, define $u_i$ such that $u_i^j=w_i^{j(i)}$.  A Lindahl equilibrium $(p,q)$ specifies a distribution over matches $q$ and a price $p_i^j$ that consumer $i$ must pay for matching $j$.  Below, we relate Walrasian equilibria of the competitive economy to the Lindahl equilibria of the collective choice market.

For any price $p$, let
$$\pi^m_i(p)=\cases
{\min_{\{j:j(i)=m\}}p_i^j & if $m\in N_i$\cr
0& otherwise
}
$$
Let $\pi_i(p)$ be the corresponding price vector and note that it assigns each partner $m\in N_i$ the price of the least cost match $j$ that yields this partner.  For any allocation $q$, define $\xi_i(q)$ such that $\xi_i^m(q)=\sum_{\{j:j(i)=m\}}q^j$.  Thus, $\xi_i(q)$ is the partner assignment implied by the allocation $q$.

\proclaim{Theorem 3}{If $(p,q)$ is a Lindahl equilibrium collective choice market, then $(\pi(p),\xi(q), q)$ is a Walrasian equilibrium of the competitive economy. If $(\pi,\xi, q)$ is a Walrasian equilibrium of the competitive economy, then $(p,q)$ such that $p_i^j=\pi_i^{j(i)}$ is a Lindahl equilibrium of the collective choice market.}

\medskip
To see why the second part of Theorem 3 is true, note that every agent's optimization problem in the collective choice market is equivalent to their corresponding optimization problem in the Walrasian setting. Condition (M3) implies that $q$ achieves the desired distribution of partners.  Moreover, since $\xi_i$ is a least cost solution for every consumer in the competitive setting,  so is $q$ in the collective choice setting.

To see why the first part of Theorem 3 is true, note that in a Lindahl equilibrium consumers choose  minimum cost solutions to their optimization problems.  Therefore, $j(i)=j'(i)$ and $p_i^j>p_i^{j'}$ implies that $q^{j}=0$.  It follows that $\xi^m_i(q)$ must be a least cost solution to the $i$'s optimization problem at prices $\pi_i(p)$.  For the same reason, the firm revenue of any matching $j$ such that $q^j>0$ is the same as in the collective choice market.  For any matching $j$ such that $q^j=0$, the firm revenue is no greater than in the collective choice market and, therefore, $q$ must be optimal for the firm.

Since we have established the existence of a Lindahl equilibrium, Theorem 3 implies that Walrasian equilibria of the matching market exist.  Corollary 1, above, shows that the Nash bargaining solution of the collective choice market can be implemented as a Lindahl equilibirum and, therefore, Theorem 3 implies that the Nash bargaining solution can be implemented as a Walrasian equilibrium of the competitive economy.

\subsection{Allocating a Finite Set of Goods and Commoditification}

In the discrete good allocation problem, there is a finite set of goods $H=\{1,\dots, r\}$ to be distributed to the $n$ agents. There is no divisible good.
Agent's $i$ utility for bundle $M$ is $v_i(M)$; we assume that $v_i(\emptyset)=0$ and $v_i(L)\leq v_i(M)$ whenever $L\subset M$. We write $v_i(h)$ instead of $v_i(\{h\})$. Let $G(H)$ denote this general set of all such utility functions. A utility function, $v_i$, is additive if $$v_i(L\cup M)=v_i(L)+v_i(M)$$ whenever $L\cap M=\emptyset$.
The discrete allocation problem can be transformed into an exchange economy by endowing each agent with one unit of fiat money and permitting random consumption.  Let $\Theta_i$ be the set of all possible random consumptions; that is, probability distributions over $2^H$ and let $$\PPP=\{\pp:2^H\rightarrow\blackboardR_+\,|\,\pp(\emptyset)=0\}$$ be the set of all prices. Hence, we allow nonadditive (or combinatorial or package) prices. The price $\pp$ is additive if $$\pp(L\cup M)=\pp(L)+\pp(M)$$ whenever $L\cap M=\emptyset$.

Then, given any price $\pp$, a consumer's budget is $$\BB(\pp):=\Big\{\theta_i\in \Theta_i\,\Big|\, \sum_M\pp(M) \theta_i(M)\leq 1\Big\}$$ and her utility maximization problem is: $$V_i=\max_{\theta_i\in\BO(\po)}\sum_{M}v_i(M)\theta_i(M)$$ A random consumption $\theta_i$ is a minimal-cost solution to the utility maximization problem if it is a solution to the maximization problem above and no other solution costs less.

Let $\AA^*$ be the set of all partitions of $H$. The firms maximization problem is: $$R(\pp)=\max_{\ao^*\in \AO^*}\sum_{M\in\ao^*}\pp(M)$$
A (feasible) allocation is $\aa=(M_1,\dots,M_n)\subset H^r$ such that $M_i\cap M_l=\emptyset$ whenever $i\not=l$; a random allocation is a probability distribution over allocations. For any allocation, $\aa=(M_1,\dots,M_n)$, $\aa_i$ denotes the $i$'th entry of $\aa$; that is, $M_i$. Let $K$ be the set of all allocations and $\Theta$ be the set of all random allocations. For any random allocation $\theta\in\Theta$, let $\theta_i$ denote the $i$'th marginal of $\theta$; that is, $$\theta_i(M)=\sum_{\ao:\ao_i=M}\theta(M)$$

The pair $(\pp,\theta)$ is a (strong) Walrasian equilibrium if for all $i$, $\theta_i$ is a minimal-cost solution to consumer $i$'s utility maximization problem at prices $\pp$ and $\sum_i(\pp(\aa_i))=R(\pp)$ for all $\aa$ such that $\theta(\aa)>0$.  It is easy to show that an arbitrary exchange economy, $v\in G^n(H)$ may have no Walrasian equilibrium in additive prices. Gul, Pesendorfer and Zhang (2020) show that every gross substitutes economy $v=(v_1,\dots, v_n)\in GS^n(H)$ has a Walrasian equilibrium with additive prices.\note{Gross substitutes utilities are a class that includes additive utilities.}   Let $\hbox{W}(v)$ be the set of Walrasian equilibrium utility vectors for $v$.

The bargaining problem associated with the discrete allocation economy $v$ is $B_v:=\coco(\{(v_1(\aa),\dots, v_n(\aa))\,|\,\aa\in K\}\cup\{o\})$.  Clearly, each allocation problem yields a unique bargaining problem;  however, the converse is not true.   Distinct allocation problems may yield the same bargaining problem.  Specifically, we may alter the property rights; that is, the specification of the goods to be traded, without affecting the bargaining problem.  In all our formulations, the entire aggregate endowment initially belongs to the fictitious firm and the $n$ agents are all endowed  with nothing other than a unit of fiat money. Nevertheless, the set of Walrasian equilibrium payoffs depends on how the goods are defined. The example below illustrates this point.

\medskip

\noindent{\bf Office allocation example:}  Three agents must decide on an office allocation.  One office will be equipped with a premium desk, one with a good desk, and one with a standard desk. We will offer two descriptions for this allocation problems that yield the same bargaining problem but different Walrasian outcomes.  The second description will yield all Lindahl equilibria as a Walrasian equilibrium outcome.

In the first version, the corresponding allocation problem has three goods and all agents have unit-demands; that is, for $M\not=\emptyset$, we have $v_i(M)=\max_{h\in M} v_i(h)$ where
$$\eqalign
{v_1(1)=10, v_1(2)=4, v_1(3)=2\cr
v_2(1)=10, v_2(2)=7, v_2(3)=3\cr
v_3(1)=10, v_3(2)=5, v_3(3)=1\cr
}
$$
Thus, good 1 is the office with the premium desk, good 2 is the office with the good desk and good 3 is the office with the standard desk.  This economy has a unique Walrasian equilibrium in which good 2 is allocated to agent $2$ and agents $1$ and $3$ each have an equal chance at getting good 1 or 3.  Hence, the equilibrium utilities are $v_1=6, v_2=7, v_3=5.5$.

Consider an alternative description of the same situation.  Each agent has a designated office; that is, each agent can derive utility only if she gets her designated office; otherwise her utility is 0. Each office is equipped with a standard desk. However, at most one desk can be upgraded to a good desk and at most one desk can be updated to a premium desk.  Hence, this specification has 5 goods, the three offices (goods 1, 2 and 3) and the two premium desks (good 4, the good desk and good 5, the premium desk).   Each agent derives utility both from her office and from any potential upgrade.  The utility functions over the relevant bundles are:
$$\eqalign
{\hat v_1(\{1,5\})=10, \hat v_1(\{1,4\})=4, \hat v_1(1)=2\cr
\hat v_2(\{2,5\})=10, \hat v_2(\{2,4\})=7, \hat v_2(2)=3\cr
\hat v_3(\{3,5\})=10, \hat v_3(\{3,4\})=5, \hat v_3(3)=1\cr
}
$$
This economy has many Walrasian equilibria and it can be shown, by appealing to Corollary 1, that the set of Walrasian equilibrium payoffs of this economy is the same its set of Lindahl equilibrium payoffs.      Notice that $B_{\hat v}=B_v$ and, thus, the two descriptions above represent the same economy in terms of achievable utility profiles. This example show that it is possible to {\it commodify} a bargaining problem $B$ in two different ways ($B_v=B_{\hat v}=B$) leading to two distinct sets of Walrasian equilibria ($W(v)\not=W(\hat v)$).

Theorem 4, below, generalizes the example. It shows that the set of Walrasian equilibrium payoffs  of any exchange economy are contained in the set of Lindahl equilibrium payoffs of the corresponding collective choice market. Moreover, every bargaining problem can be commodified (with additive utilities if $n=2$) so that the set of Walrasian equilibrium payoffs with additive prices and the set of Lindahl equilibrium payoffs are the same.

\proclaim{Theorem 4}{(i) $\hbox{W}(v)\subset \hbox{L}(v)$ for all $v\in G^n(H)$. (ii) For all $B\in\b_o$, there is $v\in G^n(H)$ such that $B_v=B$ and $\hbox{W}(v)=\hbox{L}(v)$; for $n=2$, the preceding statement holds with additive utilities and additive prices.}
\medskip

Theorem 4 and the example preceding it, reveal an advantage of Lindahl equilibria over Walrasian equilibria.  Lindahl equilibria depend only on the bargaining game while two different exchange economies that yield the same bargaining game may have two different sets of Walrasian equilibria.   At the same time,  Walrasian equilibria are simpler than Lindahl equilibria because the former involve many fewer prices.   This is so because the number of allocations typically exceeds the number of goods and because Lindahl prices are  personal while Walrasian prices are not. However, if the commodity space is rich enough, as is the commodity space we construct in the proof of Theorem 3, the distinction between Lindahl equilibrium and Walrasian equilibrium disappears.

Note that commodification does not modify the firm's technology.  Though we allow non-linear pricing, we retain the assumption of an exchange economy.  By contrast, the standard mapping used in the proof of Lemma 1 introduces a new technology that constrains the firm so that only feasible allocations can be ``produced.''

\section{Conclusion}

To achieve equitable outcomes it is often necessary to randomize.  In practice, this randomization often occurs ex ante to give priority to agents while the mechanism remains deterministic.  For example, when allocating offices, an organization may first randomly determine a priority order and then ask members to sequentially choose their preferred office.  As Hylland and Zeckhauser (1979) point out, such mechanisms lead to ex ante inefficiency.  As an alternative, they propose a market mechanism in which agents are given a budget of fiat money and choose lotteries over the available offices. The Walrasian mechanism proposed by Hylland and Zeckhouser is efficient and, if agents have identical budget, equitable but limited in its applicability.  Gul, Pesendorfer and Zhang (2019) extend Hylland and Zeckhauser's approach from unit demand to multi unit demand with gross substitutes utilities.  As we show in that paper, demand complementarities create existence problems for the standard market mechanism.  Moreover, externalities and public goods render it inefficient.  By contrast, the collective choice markets proposed in the current paper are broadly applicable to all discrete allocation problems and always efficient.

In a collective choice market, each agent expresses her demand for social alternatives rather than private outcomes.  On the one hand, this allows us to deal with a much broader range of applications but, on the other hand, the number of social alternative can be large and, therefore, the collective choice market may be too unwieldy to implement in practice. For the case of matching markets, we show that Lindahl equilibria coincide with standard competitive equilibria and, therefore, each agent need only consider the set of possible partners and not the set of possible matchings (allocations) when formulating her demand.  An important direction for future research is to examine other circumstances in which a smaller set of markets (and prices) suffices to implement Lindahl equilibria.

\section{Appendix}

\subsection{Proofs of Lemmas 1-3 and Theorem 1}

{\noindent \bf Proof of Lemma 1: } Consider the following modified economy with private goods:  let $$X=\{(x,\beta)=(x^{11},\ldots, x^{k1},\ldots, x^{1n},\ldots, x^{kn},\beta)\in \blackboardR^{kn+1}\}$$ be the commodity space.  There are $n+1$ agents and every agent has endowment zero. Agent $i$'s, for $i=1,\ldots, n$, consumption set is $$D_i=\{(x,\beta)\in X\,|\, x\ge 0, x^{jl}=0, l\not=i, \beta\ge -1\}$$ and her utility function is $U_i(x,\beta)=\sum_{j=1}^k u_i^j x^{ji}$.  Agent $n+1$ has the consumption set $$D_{n+1}=\Big\{(x,\beta)\in X\,\Big|\, \sum_{j=1}^n \min_{i} x^{ji}\ge -1, \beta\ge 0\Big\}$$  and her utility function is $U_{n+1}(x,\beta)=\beta$.

An allocation $(x_1,\beta_1,\ldots, x_{n+1},\beta_{n+1})$ is feasible if $(x_i,\beta_i)\in D_i$ for all $i$,  $\sum_{i=1}^{n+1} x_i\le 0$ and $\sum_{i=1}^{n+1} \beta_i\le 0$.  It is easy to verify that the set of feasible allocations is compact, the consumption sets are convex and bounded below, and the utility functions are quasi-concave, continuous and locally non-satiated.  Let $\tilde p$ be any price such that the utility of every agent is bounded.  Then, $\tilde p_\beta>0$ and $\tilde p^{ji}>0$ if $u^{ji}>0$.  It is straightforward to verify that at any $\tilde p$ that satisfies these conditions there exists, for every $i=1,\ldots, n+1$,  $(x_i,\beta_i), (x'_i, \beta'_i)\in D_i$ such that $\tilde p\cdot (x_i,\beta_i)<\tilde p\cdot (x'_i,\beta'_i)\le 0$.   Thus, the conditions of the equilibrium theorem (pg 12) in Gale and Mas-Colell (1975) are satisfied and the economy has a Walrasian equilibrium.
In that Walrasian equilibrium $\tilde p_\beta>0$ and, therefore, we can normalize prices so that $\tilde p_\beta=1$.  Local non-satiation of utilities implies that the equilibrium allocation is Pareto efficient and that utility maximizing consumptions are least-cost.  Pareto efficiency implies that $\sum_{j=1}^k \min_{i} x_{n+1}^{ji}= -1$.  If $x_m^{jm}+\min_{i} x_{n+1}^{ji}<0$ for some $m\in \{1,\ldots, n\}$, then utility maximization of agent $n+1$ implies $\tilde p^{jm}=0$ and, therefore, we can modify the allocation so that $x_m^{jm}=x_i^{ji}$ for all $j,m\in \{1,\ldots, n\}$ and maintain all equilibrium conditions.  Setting $q=(x^{11}_1,\ldots, x^{k1}_1), p_i=(\tilde p^{1i},\ldots, \tilde p^{ki})$ then yields the desired Lindahl equilibrium. \hfill\qed

\bigskip

\noindent{\bf Proof of Lemma 2:}  Let $x\in E(B)$.  It follows that there is a simplex $A\ge B$ such that the fair solution of $A$ is $x$.  Since $A\ge B$ it follows that $x$ is Pareto efficient and that  $d(A)\ge d(B)$ and $b(A)\ge b(B)$.  Thus,  $x=d(A)+(b(A)-d(A))/2\ge d(B)+(b(B)-d(B))/2$. For the converse, let $x\in B$ be Pareto efficient and satisfy $x\ge d(B)+(d(B)-b(B))/2$.  Since $E$ satisfies scale invariance, we may assume $d(B)=o$ and $b(B)=(1,1)$.  Since $x$ is Pareto efficient and $B\in \b$, there is $a=(a_1,a_2)>o$ such that $a\cdot x\geq a\cdot y$ for all $y\in B$. Since $x>o$, we have $ax>a_1x_1, ax>a_2x_2$.  Moreover, we must have ($2a_2x_2\geq ax\geq 2a_1x_1$) or ($2a_1x_1\geq ax\geq 2a_2x_2$).  Without loss of generality, assume the latter holds. Let $d_1=ax/a_1-2x_1$ and note that $d_1\in[0,x_1)$. Then, let $d=(d_1,0)$, $b=(a\cdot(x-d)/a_1,a\cdot(x-d)/a_2)+d$ and $A=\hbox{conv}\{d, (b_1,0), (d_1,b_2)\}$.  It is straightforward to verify that $A\geq B$ and $F(A)=x$ and, therefore, $x\in E(B)$, as desired. \hfill\qed

\bigskip

Call $A\in \b$ a {\it simplex} if $A=a\times \Delta+b$ for some $a$ such that $a_i>0$ for all $i$ and $b$ such that $b_i\ge 0$ for all $i$. We say that the simplex $A$ supports $B$ at $x$ with (exterior normal) $w$ if $x\in B\subset A$, $d(B)=d(A)$ and $wy\leq wx$ for all $y\in A$.

\proclaim{Lemma A1}{(i) For all $B$ there is a simplex $A$ that supports $B$ at $\eta(B)$ with $\nabla f_B(x)$. (ii) For any simplex $A$, $\eta(A)={1\over n}b(A)+{n-1\over n}d(A)$.}

\proof The proof of (ii) is straightforward and omitted. Assume $d(B)=o$, let $x=\eta(B)$ and let $w=\nabla f_B(\eta(B))$. Since $B$ is convex it follows that $wy\leq wx$ for all $y\in B$. Since $x_i>0$ for all $i$ it follows that $w_i=\prod_{k\not=i} x_k>0$ for all $i$.  Let ${wx/w_i}=nx_i$. Then, the simplex $A=\conv\{o,nx_1e^1,\dots, nx_ne^n\}$ supports $B$ at $\eta(B)$ with $\nabla f_B(\eta(B))$.

For arbitrary $B$, let $C=B-d(B)$. By the preceding argument,  there is a simplex $A$ that supports $C$ at $\eta(C)$ with $w=\nabla f_C(x)$. Then, note that $\eta(B)=\eta(C)+d(B)$,  $\nabla f_B(\eta(B))=\nabla f_C(\eta(C))$ and therefore, $A+d(B)$ supports $B$ at $\eta(B)$ with $\nabla f_B(\eta(B))$.\hfill\qed

\bigskip

\noindent {\bf Proof of Lemma 3: }  First, we show that $N(B)\subset E(B)$: take $x\in N(B)$. Hence, $x\in B$ and $x=\eta(A)$ for some $A$ such that $B\leq A$. By Lemma A1,  there is a simplex $C$ that supports $A$ at $x=\eta(A)$ with $\nabla_Af(\eta(A))$. Since $C$ is a simplex, $\{x\}=\{\eta(A)\}=F(C)$. Hence, $B\leq A\leq C$ and $x\in F(C)$, therefore, $x\in E(B)$.

Next, we  show that $E(B)\subset N(B)$. If $x\in E(B)$, then there is $A$ such that $\{x\}=F(A)$. Hence, $x=\eta(A)$ and $B\leq A$, therefore $x\in N(B)$.\hfill\qed

\bigskip

\noindent{\bf Proof of Theorem 1: } That $E$ satisfies scale invariance is obvious. If $A$ and $B$ are simplices such that  $B\leq A$ and $F(A)\subset B$, then $A=B$. Therefore $E(A)=F(A)$ for every simplex $A$.  Since $F(\Delta)=\{{1\over n}e\}$, it follows that $E(\Delta)=\{{1\over n}e\}$, as desired.  

To prove consistency, let $B\leq A$ and $x\in E(A)\cap B$. Then, there exists $C$ such that $A\leq C$ and $F(C)=\{x\}$. Hence, $B\leq C$ and therefore $x\in E(B)$, as desired.  To prove justifiability, assume $x\in E(B)$.  Then, there is $A$ such that $B\leq A$ and $x\in F(A)$. Hence, $A$ is a simplex and therefore $E(A)$ contains a single element $x$. That is, $E$ satisfies justifiability.

Next, assume that $S$ satisfies the axioms. We will show that $S=E$.  First, note that scale invariance and symmetry imply that $S(A)=F(A)$ for every simplex $A$.  It follows that any $x\in E(B)$ is justifiable, that is, for any $x\in E(B)$ there is $B\le A$ such that $\{x\}=S(A)$. Since $E$ satisfies scale invariance, symmetry and consistency, we conclude that $E(B)\subset S(B)$ for all $B\in \b$.  Suppose $x\in S(B)$. Then, $\{x\}= S(A)$ for some $A$ such that $B\leq A$. Since $E(A)\subset S(A)$, we must have $E(A)=\{x\}$ and since $E$ satisfies consistency, $x\in E(B)$ as desired.\hfill\qed

\subsection{Proof of Theorem 2 and Corollary 1}

Let $p$ be a price.  In a collective choice market the consumer solves
$$U_i(p)=\max_{q} u_i\cdot q \hbox{ subject to }  p_i\cdot q\le 1, e\cdot q\le 1\eqno(\hbox{P})$$
The dual of problem (P) is
$$\min_{\mu^0,\mu^1\ge 0} \mu^0+\mu^1 \hbox{ subject to } \mu^0e+\mu^1 p_i \ge u_i\eqno (\hbox{D})$$
The vector $(q,\mu_0,\mu_1)$ is feasible if $q$ satisfies the constraints of (P) and $(\mu_1,\mu_2)$ satisfies the constraints of (D).  A feasible vector $(q, \mu^0,\mu^1)$ is optimal (that is, $q$ solves (P) and $\mu^0,\mu^1$ solves (D)) if and only if
$$
\eqalign
{\mu^0(q\cdot e-1)=0\cr
\mu^1(q\cdot p_i -1)=0\cr
\hbox{ for all } j,\hskip.1in q^j(\mu^0+\mu^1p_i^j-u_i^j)=0
}
\eqno \hbox{(CS)}$$
Let $J(q)=\{j\,|\, q^j>0\}$. For any utility $u_i$ and $c_i<\max_ju_i^j$, let $\bar u_i^j(c_i)=\max\{0,u_i^j-c_i\}$. Note that $\bar u_i(c_i)$ is a utility.
\medskip

\proclaim{Lemma A2}{Let $q\in Q$ be a minimal cost solution to consumer $i$'s maximization problem for utility $u_i$ and prices $p$ such that $q\cdot p_i=1$. There are $c\geq 0$  and $\alpha>0$ such that $\alpha p_i^j\ge u^j_i-c$ for all $j$ and $\alpha p_i^j= u^j_i-c$ for $j\in J(q)$.  Moreover,  $q$ is a minimal cost solution to consumer $i$'s maximization problem for utility $\bar u_i(c)$ and prices $p$.}

\proof  Let $\mu^0,\mu^1$ be the associated solution of the dual (D). First, consider the case in which $u^j_i\neq u^m_i$ for some $j,m\in J(q)$.  Then, (CS) implies $\mu^1>0$.  Set $c=\mu^0, \alpha=\mu^1$.  Feasibility and (CS) imply that
$\alpha  p_i^j \ge u^j_i-c$ with equality if $j\in J(q)$,  as desired.

It remains to show that $q$ is a minimal cost solution to the consumer's maximization problem given utility $\bar u_i(c)$.  Since $(q,\mu^0,\mu^1)$ is feasible (for utility $u_i$), we have $\alpha p_i^j \ge u_i^j-c$ for all $j$.  We also have $\alpha p_i^j\ge \bar u_i(c)$ for all $j$, since the left-hand side is non-negative.  The first part of the Lemma implies $u^j_i-c=\bar u^j_i(c)$ for all $j\in J(q)$ and, therefore,
$q^j(\alpha p_i^j - \bar u_i(c))=0$ for all $j$.  Hence,  $(q,\mu^0,\mu^1)$ such that $\mu^0=0,\mu^1=\alpha$ is a feasible solution for (P) and (D) that satisfies (CS) for utility $\bar u(c)$; that is, $q$ solves consumer $i$'s maximization problem given utility $\bar u_i^j(c)$.  Since $u_i^j\not=u_i^m$ for some $j,m\in J(q)$, we have $\bar u_i^j(c)=u_i^j-c\not=u_i^m-c=\bar u_i^m(c)$ and therefore,  this solution must be minimal cost.

Second, consider the case where $\beta=u_i^j=u_i^m$ for all $j,m\in J(q)$.  Since $q$ is a minimal cost solution and $q\cdot p_i=1$, we must have $p_i^j=p^m_i=1$ for all $j,m\in J(q)$. Let $\mu^0,\mu^1$ be the associated solution of the dual (D).  If $\mu^1>0$, set $\alpha=\mu_1$ and repeat the argument above to conclude that $q$ is a solution maximization problem given utility $\bar u_i(c)$ and $\bar u_i^j(c)=u^j_i-c=\alpha p_i^j=\alpha>0$ for all $j\in J(q)$. Then, since $q$ is a minimal cost solution for $u_i$ implies $q$ is a minimal cost solution for $\bar u_i(c)$.

If $\mu^1=0$, set $c=\max_{\{j|p_j<1\}}u_i^j$ if $\{j\,|\,p_j<1\}\not=\emptyset$ and $c=0$ otherwise and set $\alpha=\beta-c$.  Note that $\beta-c>0$ since $q$ is a minimal cost solution. Then, $c+\alpha p_i^j\ge c\ge u_i^j $ if $p_i^j<1$ and $c+\alpha p_i^j\ge \mu^0\ge u_i^j$ if $p_i^j\ge 1$. Therefore, $\alpha p_i^j\ge u_i^j-c$ for all $j$ and with equality if $j\in J(q)$.  Then, note that $(q,c,\alpha)$ is a feasible vector that satisfies (CS) and hence $q$ is a solution to the consumer's optimization problem. Since $\beta-c>0$ and $q$ is a minimal cost solution to the optimization problem for $u_i$, it must also be a minimal cost solution for $\bar u_i(c)$.\hfill\qed
\bigskip

\proclaim{Lemma A3}{If there is $\lambda=(\lambda_1,\dots,\lambda_n)\geq o$ such that $v_i=u_i+(\lambda_i,\dots,\lambda_i)$ for all $i$, then $L(u)+\lambda \subset L(v)$.}

\proof  Suppose such a $\lambda$ exists and choose $x\in L(u)$ and let $(p,q)$ be a Lindahl equilibrium of $u$ that yields the payoff vector $x$. Then, for all $i$ and $\hat q$ such that $e\cdot \hat q\leq 1$, $u_i\cdot\hat q\leq u_i\cdot q=v_i\cdot q-\lambda_i$  and $v_i\cdot\hat q-\lambda_i\leq u_i\cdot \hat q$. It follows that $q$ is a minimal cost solution to consumer $i$'s maximization problem in $v$. Clearly, $q$ is a solution to the firm's maximization problem in the collective choice market $v$ and hence $(p,q)$ is a Lindahl equilibrium of $v$ and therefore, $x+\lambda\in L(v)$.\hfill\qed

\bigskip

\noindent{\bf Proof of Theorem 2: }  First, we show that $x\in E(B_u)$ implies $x\in L(u)$. Let $x\in E(B_u)$. Then, there is a simplex $A=na\otimes \Delta + d$ such that $B_u\leq A$ and $x=a+d\in B_u$. Assume for now, that $d=o$ so that $d(A)=d(B_u)=o$.  Then, $B_u\subset A$. For any $j\in K$, let $y^j=(u_1^j,\dots, u_n^j)$. Since $y^j\in B_u\subset A$, it is a unique convex combination of the extreme points of $A$. Let $z^j_i$ be the weight of $na_ie^i$ in that convex combination and set $p_i^j=nz^j_i$.  Note that $z_i^j>0$ if and only if $p_i^j>0$ and that for all $j\in K$ such that $p_i^j>0$, we have $u_i^j/p_i^j=a_i$.
Since $x\in B_u$, it is a convex combination of the extreme points of $B_u$. Let $q^j$ be the weight of $y^j$ in that convex combination.  By construction, $u_i\cdot q=a_i$.  Note that for any $\hat q$ such that $\hat q\cdot p_i\leq 1$ we have $$u_i\cdot\hat q=\sum_{j=1}^K a_i p_i^j\cdot \hat q^j\le a_i$$
with equality only if $\hat q\cdot p_i=1$. Therefore, $q$ is a minimal cost solution to the consumer's problem and $(p,q)$ is a Lindahl equilibrium.

If $o=d(B_u)\not=d$; that is, $d(B_u)\leq d(A)$, then repeat the above construction for $A'=A-d$ to show that $x-d$ is a Lindahl equilibrium payoff vector of the economy $v$ such that  $v_i^j=u_i^j-d_i$. Then, Lemma A3 ensures that $x$ is a Lindahl equilibrium payoff vector for the economy $u$.

Next, we show that $x\in L(u)$ implies $x\in E(B_u)$.  Let $x\in L(u)$ and let $(p,q)$ be the corresponding Lindahl equilibrium outcome.  First, consider the case in which $p_i \cdot q=1$ for all $i$.  By Lemma A2, there is, for each $i$, some $c_i\geq 0$ and  $\alpha_i>0$ such that  $u_i^j-c_i\ge \alpha_i p_i^j$ for all $j$ with equality for $j\in J(q)$.  Furthermore, Lemma A2 implies that $(p,q)$ is also a Lindahl equilibrium for the collective choice market $(\bar u_i(c_i))_{i=1}^n$.  Since $\bar u_i^j(c_i)=\alpha_i p_i^j$ for all $j\in J(q)$ and $p_i\cdot q=1$, we have $\bar u_i(c_i) \cdot q=\alpha_i p_i\cdot q=\alpha_i$ and $x_i=c_i+\alpha_i$.  Let $z\in B_u$ and let $\hat q$ be such that $z_i=u_i\cdot\hat q$.

Since $\alpha_i p_i^j \geq u_i^j-c_i$ for all $i,j$, we have $\alpha_i p_i\cdot\hat q\ge z_i-c_i$ for all $i$ or $p_i\cdot\hat q\ge (z_i-c)/\alpha_i$.  Note that the firm's profit cannot be greater than $n$, the aggregate endowment of money.  Therefore, firm optimality implies $n\geq\sum_{i=1}^n p_i\cdot q\geq\sum_{i=1}^n p_i\cdot\hat q$.   Thus, we obtain
$$n\ge \sum_{i=1}^n {z_i-c\over \alpha_i}\eqno(1) $$
for all $z_i\in B_u$.
Let $d=(c_1,\ldots, c_n)$, $a=(\alpha_1,\ldots, \alpha_n)$ and $A=na\otimes \Delta +d$ and note that (1) implies $A\ge B_u$.  Clearly, $F(A)=a+d=x$ and, therefore, $x\in E(B_u)$.

Finally, consider the case in which $p_i\cdot q<1$ for some $i$. Let $I$ be the set of all such agents. Let $J^*_i=\{j|u_i^j\ge u_i^m\,\forall m\}$ be the bliss outcomes for $i$.  If $p_i\cdot q<1$ for $i\in I$, then $J(q)\subset J^*_i$ since otherwise $i$ is not choosing a utility maximizing plan.  Furthermore, since $q$ is a minimal cost solution for consumer $i\in I$, $p_i^j=p_i^m$ for all $i\in I$ and $j,m\in J_i(q)$. Define $\bar p=(\bar p_1,\ldots, \bar p_n)$  as follows: $\bar p_i^j=1$ if $i\in I$ and $j\in J_i$ and $\bar p_i^j=p_i^j$ otherwise.  It is easy to see that $(\bar p,q)$ is a Lindahl equilibrium.  Moreover, every consumer satisfies $\bar p_i \cdot q=1$ for all $i$.  Thus, we can apply the argument above to show that $x\in E(B_u)$. \hfill\qed

\medskip

{\bf Proof of Corollary 1:  }  For $c\in C(u)$, define $H_i(c_i)=\hbox{conv}\{e^j|\, u_i^j\ge c_i,j\in K\}$ and note that $\bar u_i^j(c)\cdot q+c=u_i^j\cdot q$ if and only if $q\in H_i(c_i)$.  Let $H(c)=\bigcap_{i=1}^n H_i(c_i)$. Let $q$ be the Nash allocation for $\bar u(c)$.  Note that $c$ is admissible if and only if $q\in H(c)$.

To prove the first statement of the corollary, assume $c$ is admissible and $q$ is a Nash allocation for $\bar u(c)$. Hence, $q\in H(c)$ and for all $i$, $\bar u_i^j(c_i)>0$ for some $j$ and, therefore, the constraint $q\cdot e\le 1$ is binding.   Then, the first order condition of (N) implies that there is $\lambda>0$ such that
$$\sum_{i=1}^n {\bar u_i^j(c_i)\over \bar u_i(c_i)\cdot q}\le \lambda \eqno(F)$$
and
$$\sum_{i=1}^n {\bar u_i^j(c_i)q^j\over \bar u_i(c_i)\cdot q}= \lambda q^j$$
Then, $q\cdot e=1$ implies $\lambda=n$.  Set $\mu_i^0=c_i, \mu_i^1= \bar u_i(c_i)\cdot q$ and $p_i=\bar u_i(c_i)/(\bar u_i(c_i)\cdot q)$.  Hence, $p_i\cdot q=1$ and $q\in H_i(c_i)$ and therefore,
$\mu_i^0+\mu_i^1 p_i^j\ge u_i^j$ for all $j$.  Moreover, equality holds for all $j$ such that $q^j>0$.  That is, $(\mu^i_0, \mu^i_1, q)$ is a feasible solution that satisfies (CS) and, therefore, is optimal. It remains to show that $q$ solves the firm's problem at prices $(p_1,\ldots, p_n)$.   Since $\lambda=n$, inequality (F) implies that $\sum_{i=1}^n p_i^j e_i^j\le n$ for all $j$ and, therefore, firm profit cannot exceed $n$.  Since $q$ yields profit $n$, it maximizes profit.

To prove the second statement of the corollary, assume that $(p,q)$ is a Lindahl equilibrium.  Then, consumer optimality implies that there are $\mu_0^i, \mu_1^i$ such that $\mu_i^0+\mu_i^1 p_i^j\ge u_i^j$ for all $j$ where equality holds for all $j$ such that $q^j>0$. Therefore, $\mu_i^0+\mu_i^1=u_i\cdot q$ and $\mu_i^1\ge \bar u_i^j(\mu_0^i)$ for all $j$ with equality if $q^j>0$.  Summing over all $i$ we obtain
 $$\sum_{i=1}^n {\bar u_i^j(c_i)\over \mu_i}=\sum_{i=1}^n {\bar u_i^j(c_i)\over \bar u_i(\mu_0^i)\cdot q}\le n \eqno(F)$$
 with equality if $q^j>0$.  Thus, $q$ is the Nash allocation for $\bar u(c)$.  Moreover, if $e^j\not\in H_i(\mu_i^0)$ then $\mu_0^i+\mu_1^i p_i^j\ge \mu_0^i>u_i^j$ and, therefore, $q^j=0$.  Hence $q\in H_i(\mu_i^0)=H_i(c_i)$ for all $i$.  \hfill\qed

\subsection{Proof of Theorems 3 and 4}

{\noindent \bf Proof of Theorem 3: }Lemma A1(i) yields a simplex $A=\conv\{z_1e^1,\dots, z_ne^n\}$ ($z_i>0$ for all $i$) that supports $B_w$ at $\eta(B_w)$ with $\nabla f_{B_w}(\eta(B_w))$.  Hence, $\eta(A)=\eta(B_w)$ and $d(B)=d(A)=o$ and therefore, by Lemma A1(ii), $\eta(B_w)={1\over n}e\otimes z$. Let $q$ be any probability distribution over $K$ such that $\sum_jq^j(w_1^j,\dots, w_n^j)=\eta(B_w)$.

Each $x\in B_w$ and is a unique convex combination of the extreme point of $A$. That is, $x_i=\lambda_i\cdot z_i$ for $\lambda_1,\dots,\lambda_n\geq0$ such that $\sum_i\lambda_i\leq 1$. Let $\lambda_i(x)$ be the weight of $z_i$ in that convex combination. For $x=(w_1^j,\dots, w_n^j)$, let $p_i^j=\lambda_i(w_i^j)$ and let $\pi_i^m=p_i^j$ for some $j$ such that $j(i)=m$. Clearly, $\lambda(w_i^j)=\lambda(w_i^{j'})$ whenever $j(i)=j'(i)$. Hence, $\pi_i^m$ is well-defined. Verifying that $(\pi,q)$ is a WE and $(p,q)$ is a LE is straightforward.

To conclude the proof, we will show that any WE is an LE. Let $(\pi,q)$ be an WE and let $p_i^j=\pi_i^{j(m)}$. Then, clearly, $(p,q)$ is a WE.\hfill\qed

\bigskip
{\noindent \bf Proof of Theorem 4: }Let $(\pp,\theta)$ be a WE and let $K=\{\aa(1),\dots, \aa(k)\}$ be the set of all feasible allocations. Then, for every feasible allocation $\aa(j)$, let $$p_i^{j}=\sum_{h\in \ao_i}\pp^h$$ Then, it is easy to verify that $(p,\theta)$ is an LE.

To prove the second assertion, let $n=2$ and $y^1=(y^1_1,y_2^1),\dots, y^k=(y^k_1,y_2^k)$ be all of the vertices on the Pareto frontier of $B\in\b_o$ ordered so that $y^1_1<y^2_1<\dots<y^k_1$.
First consider the case in which  $y^1_1=0$ and $y^k_2=0$.
Then, let $H=\{1,\dots, k-1\}$, $v_1(h)=y^{h+1}_1-y^{h}_1$ and
$v_2(h)=y^{h}_2-y^{h+1}_2$ for $h\in H$. For $\emptyset\not=M\subset H$, let $v_i(M)=\sum_{h\in M}u_i(h)$ for $i=1,2$. Hence, $v_i$ is additive.

Since $f(h)={y^{h}_2-y^{h+1}_2\over y^{h+1}_1-y^{h}_1}$ is an increasing function, in any efficient allocation of the economy $v$, there is $l$ such that agent 2 consumes $M_1(l)=\{1,\dots, l\}$ and agent 1 consumes $M_2(l)= \{l+1,\dots k-1\}$.  Hence, $B_{v}=B$. A random allocation is Pareto efficient if and only if it has at most two elements in its support, and hence is of the form $\aa(l)=(\aa_1(l),\aa_2(l))=(M_1(l),M_2(l))$ with probability $\gamma\in (0,1]$ and $\aa(l+1)=(\aa_1(l+1),\aa_2(l+1))=(M_1(l+1),M_2(l+1))$ with probability $1-\gamma$.

Let $x\in B_v$ be a Lindahl equilibrium payoff vector.  Hence, $q^{l}$ the probability of the allocation $\aa(l)$, is $\gamma$ and $q^{l+1}=1-\gamma$ for $\aa(l)$ and $\aa(l+1)$ as above. By Lemma 2, $x\geq b(B_v)/2$; in particular, $x_1\geq y_1^k/2$. Since $x$ is Pareto efficient, there is $a\in\blackboardR^2_{++}$ such that $a x\geq az$ for all $z\in B_v$. We must have $a_2x_2\geq a_1x_1$ or $a_1x_1\geq a_2x_2$.  Without loss of generality, assume the latter holds.  Then, let $\pp^{h}={v_1(h)/x_1}$ for all $h\leq l$ if $\gamma=1$ and for all $h\leq l+1$ (if $\gamma<1$). Hence, the random consumption that yields $\aa_1(l)$ with probability $\gamma$ and $\aa_1(l+1)$ with probability $1-\gamma$ is just affordable for agent 1.  Moreover, agent 1's util per price of the goods she is to consume is  constant at $1/x_1$.

Note that since $x_1\geq y_1^k/2$, if we were to set the price, $\pp^h$, of each of the remaining goods to ${v_1(h)/x_1}$, agent 2 could purchase the random consumption that yields $\aa_2(l)$ with probability $\gamma$ and $\aa_2(l+1)$ for less than 1, her endowment of fiat money.

Conversely, if we were to set $\pp^h$ for each of the remaining goods to ${a_2v_2(h)/a_1x_1}$, agent 2 would not be able to afford the random consumption that yields $\aa_2(l)$ with probability $\gamma$ and $\aa_2(l+1)$  if $a_2x_2>a_1x_1$. Let $I^h=\Big[{v_1(h)\over x_1},{a_2v_2(h)\over a_1x_1}\Big]$.  Since $a_2/a_1$ is the slope of the tangent line through the point $x$, $I^h\not=\emptyset$ for all $h> l+1$ if $\gamma<0$ and for all $h>l$ if $\gamma=1$.  Note also that if $\gamma<1$, the interval $I^{l+1}$ is a single point. Hence, we can choose $\pp^h\in I^h$ for all such $h\geq l$ of $\gamma<1$ and for all $h\geq l+1$ if $\gamma=1$ so that agent 2 can just afford the random consumption $(\gamma, \aa_2(l); 1-\gamma, \aa_2(l+1))$. Moreover, agent 2's util per price of the goods he consumes in no greater than $a_2/a_1x_1$.

Since the slope of $f$ is increasing, agent 1's util per price for goods $h\geq l+1$ is greater than $1/x_1$ while agent 2's util per price of goods $h\leq l$ is greater than $a_2v_2/a_1x_1$. It follows that $(\gamma, \aa_i(l); 1-\gamma, \aa_i(l+1))$ is a minimal cost solution to the maximization problem of agent $i$. Hence, $x\in W(v)$.

If $y^{k-1}_2>0$, then let $H=\{1,\dots, k\}$ and  define the utility of the new good $k$ as follows: $ v_1(k)=0$ and $v_2(k)=y^{k-1}_2$. If $y_1^1>0$, then add a new good $0$, (or another new good $0$, if $k$ has already been added) to $H$ such that $v_1(0)=y_1^1$ and $v_2(0)=0$.  Then, repeat the construction above, after extending each  $v_i$ to all subsets of $H$ additively to derive, once again, a WE that yields $x$ as its payoffs.

For the general case; that is, for $n\geq 2$ and any polytope $B\in\b_o$, we define the commodity space as follows: let $X$ be the set of Pareto efficient extreme points of $B$. Then, let $X_i$ be the projection of $X$ to the $i$'the coordinate; that is, $X_i=\{x_i\,|\, x\in X\}\backslash\{0\}$ and define the set  of possible partial payoff vectors as
$$X^0=(X_1\cup\{-1\})\times(X_2\cup\{-1\})\times\cdots\times(X_n\cup\{-1\})$$
For $y\in X^0, x_i\in X_i$ let $(y_{-i},x_i)\in X^0$ be the vector in which the $i-$coordinate of $y$ is replaced by $x_i$.
Define the set of consistent partial payoff vectors and minimally inconsistent partial payoff  (MIPP) vectors as follows:
$$\eqalign{ X^+&=\{x\in X^0\,|\, \hbox{ there is }y\in X \hbox{ such that } x_i\in\{-1,y_i\}\hbox { for all $i$ }\}\cr X^-&=\{x\in X^0\backslash X^+\,|\hbox{ there is $i$ such that } (x_{-i},-1)\in X^+\}}$$ For any $x\in X^-$, let $\delta(x)$ be the cardinality of the set $\{i\,|\,x_i\not=-1\}$ minus $1$ and let $Y(x)$ be a set that contains exactly $\delta(x)$ copies of $x$.
 Define the set of goods as follows: $$H=\bigcup_iX_i\cup\bigcup_{x\in X^-}Y(x)$$ Hence, $H$ contains one copy, $y_i\in X_i$, of every possible payoff for each agent and $\delta(x)$ copies of every MIPP  $x\in X^-$; that is, one fewer copy than the number of possible payoffs in the MIPP vector $x$ (i.e., entries that are not $-1$). Then, for any $M\subset H$, define the utility of agent $i$ for the bundle $M$ as follows: for all $y_i\in X_i$, $g_i(M,y_i)=0$ if there is $\hat y_i\leq y_i$ such that $\hat y_i\in X_i\backslash M$ or there is $x\in X^-$ such that $x_i=y_i$ and $Y(x)\cap M=\emptyset$; otherwise  $g_i(M,y_i)=1$. Then $$v_i(M)=\max\{y_i\,|\, g_i(M,y_i)=1\}$$

Next, we will show that $B_v=B$. Take $y\in X$ and for every $i$ and $x\in X^-$ such that $x_i=y_i$, choose a distinct $x^{l_i}\in Y(x)$. Then, let $M_i$ be the set consisting of each such $x_i^{l_i}$ and all $\hat y_i\leq y_i$ in $X_i$. Clearly, $v({M_i})=y_i$. To see that this collection of $M_i$'s, for $i=1,\dots, n$, is feasible, note that if such a collection were not feasible, there would be some $x\in X^-$ such that the cardinality of the set $\{i\,|\,x_i=y_i\}$ is greater than the cardinality of the set $Y(x)$. But this contradicts the definition of $\delta(x)$.

Thus we have shown $X\subset B_v$ and therefore $\coco(X\cup\{o\})=B\subset B_v$.  To see that $B_v\subset B$, take any collection of disjoint subsets $M_1,\dots M_n,\subset H$. Let $y_i=v_i({M_i})$ for all $i$ and assume that $y=(y_1,\dots, y_n)\notin B$.  Hence, there is $x\in X^-$ such that $x_i=y_i$ for all $i$ such that $x_i\not=-1$.  Then, each $M_i$ must contain an element of $Y(x)$, contradicting the definition of $\delta(x)$.

Finally, we will show that $L(v)\subset W(v)$. Let $K$ be the set of all possible allocations of goods for the competitive economy $v$. For any allocation $\aa\in K$, let $u^{\ao}_i=v_i(\aa_i)$.
Take any LE random outcome $q$.
By Corollary 1, there is an LE price vector $p$ for $q$ such that $p^j_i=p^l_i$ whenever consumer $i$ receives the same bundle in allocation $j$ and $l$. Hence, for any $M\subset H$,  let $\pp(M)=p_i^{\ao}$ if $M$ is a minimal subset of $\aa_i$ such that $v_i(M)=v_i(\aa_i)$. For all remaining $M\subset H$, let $\pp(M)$ be the maximum of $\pp(L)$ among all $L\subset M$ for which $\pp(L)$ has been defined above.

Then, let $\theta_i(q)$ be the random consumption  of player $i$ implied by the random outcome (i.e., random allocation) $q$ and let $\theta(q)=(\theta_1(q),\cdots\theta_n(q))$.
Note that consumer $i$ gets the same utility with any $\hat q$ in the collective choice problem as she gets with consumption $\theta_i$ and the two cost the same. Moreover, for every random consumption $\hat\theta_i$ in the competitive economy, there is a random outcome $\hat q$ that yields the same utility and costs the same. Then, the minimal-cost optimality of $\theta_i$ follows from the minimal cost optimality of $q$.

Suppose there is some collection $\{M^1,\dots, M^l\}\in\AAA^*$ such that $\sum_{k=1}^k\pp(M^l)>R(\pp)$. First, we note that from the definition of $v_i$ it follows that $k\not=m$ implies $v_i(M^k)\cdot v_i(M^m)=0$ for all $i$. Hence, there is a collection $\{M_1,\dots M_n\}\subset \{M^1,\dots, M^l\}\cup\{\emptyset\}$ such that for any $M^k$ not in this collection, $v_i(M^k)=0$ for all $i$. Hence, $\pp(M^k)=0$ for all such $M^k$, $\aa:=(M^1,\dots, M^l)\in\AAA$ and $\sum_i\pp(M_i)>R(\pp)$. But, $\pp(M_i)=p^{\ao}_i$ for all $i$, contradicting the firm-optimality of $(p,q)$ in the Lindahl equilibrium.
\hfill\qed

\nextpage

\frenchspacing \singlespace
\parskip=8pt plus1pt minus1pt
\parindent=0pt
\centerline{\twelvebf References}
\bigskip






Fujishige, S. and Z. Yang (2003) ``A Note on Kelso and Crawford's Gross Substitutes
Condition,'' {\sl Mathematics of Operations Research,} 28, 463--469.

Foley, D. (1967) Resource Allocation and the Public Sector, {\sl Yale Economic Essays}, 3 (Spring, 1967)



Hyland, A. and R. J. Zeckhauser (1979)  ``The Efficient Allocation of Individuals to Positions," {\sl Journal of Political Economy,} 87, 293--314.

Kalai, E. and M. Smorodinsky (1975)  ``Other Solutions to Nash's Bargaining Problem,'' {\sl Econometrica},
43, 513--518.


Kelso, A. S. and V. P. Crawford (1982) ``Job Matching, Coalition Formation, and Gross
Substitutes," {\sl Econometrica,} 50, 1483--504.


Lindahl, E. (1919)  Just Taxation - A Positive Solution.  Reprinted in part in: R.A. Musgrave and A.
Peacock, eds., Classics in the theory of public finance (Macmillan, London, 1958).

Mas-Colell, A., (1992) ``Equilibrium Theory with Possibly Satiated Preferences,'' Equilibrium
and Dynamics: Essays in Honour of David Gale, edited by M. Majumdar,
London, Macmillan, 201--213.


Thomson, W, (1994) ``Cooperative Models of Bargaining," in: Handbook of Game Theory, Volume 2, Edited by R.J. Aumann and S. Hart, Elsevier Science B.V., 1238--1284.



Nash, J.F. (1950) ``The Bargaining Problem,'' {\sl Econometrica}, 28: 155--162.



Perles, M.A. and M. Maschler (1981) ``A Super-Additive Solution for the Nash Bargaining Game,''
{\sl International Journal of Game Theory}, 10: 163-193.

D. Schmeidler and K. Vind,  (1972), ``Fair Net Trades,'' {\sl Econometrica} 40, 637--642.


Hal R. Varian (1973) ``Equity, Envy, and Efficiency" {\sl Journal of Economic Theory}, 9, 63--91



\end
Hajek, Bruce (2008), ?Substitute valuations: Generation and structure.? Performance
Evaluation, 65, 789?803. [854]
Hatfield, John W., Nicole Immorlica, and Scott D. Kominers (2012), ?Testing substitutability.?
Games and Economic Behavior, 75, 639?645. [854]
Hatfield, JohnW., ScottD. Kominers, AlexandruNichifor,MichaelOstrovsky, and AlexanderWestkamp
(2013), ?Stability and competitive equilibrium in trading networks.? Journal
of Political Economy, 121, 966?1005. [854]
Hatfield, JohnW. and Paul R.Milgrom (2005), ?Matching with contracts.? American Economic
Review, 95, 913?935. [853, 854, 855, 856, 859, 861]

Lehmann, Benny, Daniel Lehmann, and Noam Nisan (2006), ?Combinatorial auctions
with decreasing marginal utilities.? Games Econom. Behav. 55, 270?296. [860]
[859, 860]
Ostrovsky,Michael (2008), ?Stability in supply chain networks.? American Economic Review,
98, 897?923. [854]
 [859]
Paes Leme, Renato (2014), ?Gross substitutability: An algorithmic survey.? Unpublished
paper, Google Research NYC. [862]
Reijnierse, Hans, Anita van Gellekom, and Jos A. M. Potters (2002), ?Verifying gross substitutability.?
Economic Theory, 20, 767?776. [854]
Roth, Alvin E. (1984), ?Stability and polarization of interests in job matching.? Econometrica,
52, 47?58. [853]
Shapley, Lloyd S. (1962), ?Complements and substitutes in the optimal assignment problem.?
Naval Research Logistics Quarterly, 9, 45?48. [855]
. [853, 854

\end